\documentclass{jfm}

\usepackage{graphicx}
\usepackage{newtxtext}
\usepackage{newtxmath}
\usepackage{natbib}
\usepackage{hyperref}
\hypersetup{
    colorlinks = true,
    urlcolor   = blue,
    citecolor  = black,
}

\newcommand{\RomanNumeralCaps}[1]
\linenumbers

\usepackage{mathtools}
\usepackage{amsmath}
\usepackage{amssymb}

\title{A physics-inspired alternative to spatial filtering for large-eddy simulations of turbulent flows}

\author{Perry L. Johnson%\aff{1}
  \corresp{\email{perry.johnson@uci.edu}}}

\affiliation{Department of Mechanical and Aerospace Engineering, University of California, Irvine, CA, USA}

\begin{document}
\maketitle

% < 250 words for JFM
\begin{abstract}
Large-eddy simulations (LES) are widely-used for computing high Reynolds number turbulent flows. Spatial filtering theory for LES is not without its shortcomings, including how to define filtering for wall-bounded flows, commutation errors for non-uniform filters, and extensibility to flows with additional complexity, such as multiphase flows. In this paper, the theory for LES is reimagined using a coarsening procedure that imitates nature. This physics-inspired coarsening (PIC) approach is equivalent to Gaussian filtering for single-phase wall-free flows but opens up new insights for modeling even in that simple case. Boundaries and nonuniform resolution can be treated seemlessly in this framework without commutation errors, and the divergence-free condition is retained for incompressible flows. An alternative to the Germano identity is introduced and used to define a dynamic procedure without the need for a test filter. Potential extensions to more complex physics are briefly discussed.
\end{abstract}

%\begin{keywords}
%Authors should not enter keywords on the manuscript, as these must be chosen by the author during the online submission process and will then be added during the typesetting process (see \href{https://www.cambridge.org/core/journals/journal-of-fluid-mechanics/information/list-of-keywords}{Keyword PDF} for the full list).  Other classifications will be added at the same time.
%\end{keywords}

%{\bf MSC Codes }  {\it(Optional)} Please enter your MSC Codes here

\section{Introduction}
\label{sec:intro}

Because of the inherently multiscale nature of turbulence, the fine grids and small time steps required for direct numerical simulation (DNS) of turbulent flows are prohibitively expensive for many higher Reynolds number flow applications.
Large eddy simulation (LES) has become a popular technique for computing and predicting turbulent flows. In LES, the computational grid may be significantly coarser than the requirements of DNS, directly resolving only the large-scale features of the flow. Sub-grid models are introduced to approximate the net effect of unresolved fluctuations on the large-scale dynamics.

The theory for LES is based on reducing the number of computational degrees of freedom (DoFs) necessary for a digital representation of a turbulent flow, i.e. increasing the required grid spacing, typically using a low-pass spatial filter \citep{Leonard1975, Germano1992, Sagaut2006, Moser2021}.
The practical significance of spatial filtering theory for LES is not necessarily obvious, because most approaches assume an implicit filter.
That is, an explicit calculation of a spatial filter is not necessary or even commonly employed in LES practice. What the theory of spatial filtering does provide is a definition of what the simulation aims to reproduce, and more importantly, a partial differential equation for the evolution of the resolved flow features, including a mathematical expression for the unclosed term(s) that need to be modeled. The definitions provided by spatial filtering enable the direct calculation of unclosed terms from DNS data, allowing for the \textit{a priori} testing of candidate models \citep{Borue1998} as well as the theoretical derivation or justification of many proposed models \citep{Clark1979, Bardina1980, Verstappen2011, Rozema2015}.
Perhaps the most notable fruit of spatial filtering theory for LES practice is the widely-used dynamic procedure for determining model coefficients on the fly using an explicit test filter \citep{Moin1991, Germano1991, Lilly1992, Vreman1994a, Meneveau1996, Bou-Zeid2005}.
Dynamic models are based on the Germano identity, which relates fields filtered at two different filter widths \citep{Germano1992}.

However, spatial filtering LES theory is not without drawbacks, which have motivated some proposed modifications and alternative formulations. % \citep{Langford1999, Fox2003, Pope2010}.
The optimal LES approach introduced by \citet{Langford1999} addresses the inherent loss of information by invoking conditional averages and seeking a formulation for predicting single-time, multi-point statistics while minimizing short-time errors in general. Ultimately, it still relies on spatial filtering and assumes that no history information may be used for modeling, e.g., by Lagrangian averaging \citep{Meneveau1996, Bou-Zeid2005} or evolving a sub-grid kinetic energy equation \citep{Kim1996}.

The self-conditioned fields approach introduced by \citet{Fox2003} and \citet{Pope2010} redefines the objective of LES as the (unfiltered) velocity conditioned on the filtered velocity field (or another lower dimensional representation). This conception of LES has many positive characteristics but makes direct (\textit{a priori}) testing of models practically impossible. Self-conditioned fields LES also introduces additional closure terms pertaining to gradients with respect to conditioning variables (for which no models are proposed beyond ignoring the terms for expediency), and it largely does not provide much insight into how to construct models, mostly serving to justify existing models.

One significant difficulty with spatial filtering theory for LES is that nonuniform spatial filters do not commute with spatial differentiation, leading to commutation errors and loss of divergence-free velocity fields in the case of incompressible flows \citep{Ghosal1995, Langford2001, Yalla2021}.
In practice, commutation errors are typically neglected and a divergence-free condition is imposed without justification.
Another difficulty with spatial filtering arises close to a boundary, where the filter operator requires flow information from outside the fluid domain \citep{Drivas2018}, unless a nonuniform filter size is used \citep{Bose2014} leading to the commutation errors described above. The filtered field may also be conceived of as the solution to an elliptic partial differential equation \citep{Germano1986a, Germano1986b, Bull2016}, in which case the choice of boundary conditions may alleviate this particular difficulty \citep{Bae2017}. However, common LES treatments near the wall simply revert to RANS-based techniques without full justification in terms of spatial filtering. This is true for hybrid RANS-LES models \citep{Piomelli2002, Frohlich2008, Spalart2009, Mockett2012} as well as wall-modeled LES \citep{Larsson2016, Bose2018}.
The self-conditioned fields approach does manage to circumvent commutator errors altogether and to provide a well-defined near-wall behavior matching the no-slip boundary condition \citep{Pope2010}, though these positive characteristics are counter-balanced by the challenges noted in the previous paragraph.

With interest growing in performing LES-like calculations with increasing physical complexity, spatial filtering theory is encountering more challenges. In multiphase flows, for instance, direct application of spatial filtering blurs the interface \citep{Labourasse2007}, unnecessarily losing its distinctiveness even though sharp interface methods exist for careful, robust treatment such discontinuities.
It is desirable to have an LES theory that allows for retaining sharp interfaces while removing features such as small-scale ripples that cannot be resolved on a coarse grid \citep{Tryggvason2020}.
One alternative is the dual-scale approach of \citet{Herrmann2013}, but this requires DNS-like resolution for the phase indicator field (or volume fraction field) and a super-resolution enrichment for the velocity \citep{Herrmann2018}.

In the simulation of turbulent flows laden with small particles, including unresolved droplets or bubbles in multiphase flows, it is common to avoid the need to resolve the flow around a particle (i.e., computationally resolve the proper boundary or interface conditions at the particle surface) using a Lagrangian tracking approach. A closure such as a drag law must be introduced. When the particle is much smaller than the Kolmogorov scale, the Reynolds number based on its relative velocity is small, and the particle is not near a flow boundary, the Lagrangian tracking approach can be quite precise \citep{Maxey1983, Balachandar2010}. In other more complex scenarios, Lagrangian tracking represents an approximate treatment motivated by computational tractability not unlike spatial filtering for LES. In fact, if the goal of reducing computational DoF is indeed central to LES, then the mathematical theory of LES should be fundamentally compatible with an Eulerian-Lagrangian approach to particle-laden flows in a way that spatial filtering is not. This includes potential hybrid approaches to multiphase flows that combine the direct resolution of large-scale interface features with a Lagrangian tracking approach for small-scale features \citep{Kim2020}.
%For unresolved interfaces, it would also be beneficial to develop a theory of LES that would lead to a point-particle representation for particle-laden flows in the proper limits.
%Similarly, for small particles, drops and bubbles, Lagrangian tracking provides a particularly efficient method for representing multiscale phenomena with a reduced number of computational degrees of freedom. Leveraging physics-based and data-driven models for single-particle and multi-particle drag laws and correction schemes (HORWITZ,BALA) in the context of under-resolved simulations (LES) would benefit from a framework that can go beyond simply spatial filtering.

While LES may be practiced in isolation from specific concerns of a consistent framework, a specific definition of that which an LES aspires to accurately reproduce is required for advanced techniques such as data-driven closure \citep{Sarghini2003, Moreau2006, Vollant2017, Gamahara2017, Wang2018, Yang2019, Zhou2019, Cheng2019, Beck2019, Sirignano2020, Xie2020a, Xie2020b, Yuan2020, Park2021, Bode2021, Freund2021, Stoffer2021, Portwood2021, Wang2021, Prakash2021, Duraisamy2021}
and super-resolution enrichment \citep{Scotti1999, Stolz1999, Domaradzki1999, Milano2002, Leonard2016, Ghate2017, Maulik2017, Bassenne2019, Wang2019, Ghate2020, Liu2020, Kim2021}.
For example, without a clear definition of what an LES solution should represent, one cannot train a neural network to serve as a sub-grid closure in a robust way. In this sense, the particular relationship between an LES solution and the fully-resolved flow must be afforded more direct attention in the age of data. This is especially true for more complex scenarios with solid boundaries, interfaces, and multiphysics; where pure spatial filtering is not likely the optimal approach.

The scope of this paper is as follows. This paper introduces the concept of physics-inspired coarsening (PIC) as a proposed alternative to spatial filtering as a basis for LES.
In addition to laying the groundwork of PIC theory, results for the simplest of turbulent flows are demonstrated with \textit{a priori} and \textit{a posteriori} testing of representative PIC-based models. Beyond that, the paper explores the extensibility of PIC theory for complex flows to motivate future work in various directions. The basic theory of PIC is established in \S \ref{sec:PIC}. Following that, \S \ref{sec:wall-free} explores in detail single-phase unbounded flows with uniform resolution, for which PIC is equivalent to Gaussian filtering. Even with this equivalence, PIC theory provides new insights into energy cascade physics and model development. Preliminary models based on these insights are demonstrated with \textit{a priori} and \textit{a posteriori} testing. In \S \ref{sec:complex}, PIC theory is expanded to include a number of more complex effects such as anisotropic and nonuniform resolution, heat and mass transfer, and flows with solid boundaries and multiphase interfaces. Conclusions are drawn in \S \ref{sec:conclusions}.

%For a general discussion of spatial filtering and LES, the reader is referred to review articles by \citet{Meneveau2000, Moser2021}.

% NOT: difficulties of relying on a priori analysis, implicit approaches work for simple flows because of specific physics of turbulent energy cascade, lack a clear path to generalization

\section{Physics-inspired coarsening (PIC)}\label{sec:PIC}

The underlying philosophy of PIC is to view the removal of DoFs from a flow more as an imitation of nature than as an image processing trick.
As a primary example, the viscosity, $\mu$, provides a natural mechanism that prevents the creation of motions smaller than the Kolmogorov microscale, $\eta = \nu^{3/4} \epsilon^{-1/4}$, where $\nu = \mu/\rho$ is the kinematic viscosity, $\rho$ is the fluid mass density, and $\epsilon$ is the turbulent dissipation rate. 
This natural mechanism may be imitated for LES theory by conjuring an artificial viscous process to further smooth a frozen snapshot of turbulence, removing small motions below a cutoff length scale, $\ell > \eta$.  The artificial physics-inspired process evolves in a pseudo-time that is independent of physical time. The overall idea of physics-inspired coarsening is illustrated in Figure \ref{fig:schematic} and described in detail in the remainder of this section.

\subsection{Navier-Stokes equation}
\begin{figure}
	\centering
	\includegraphics[width=1.0\linewidth]{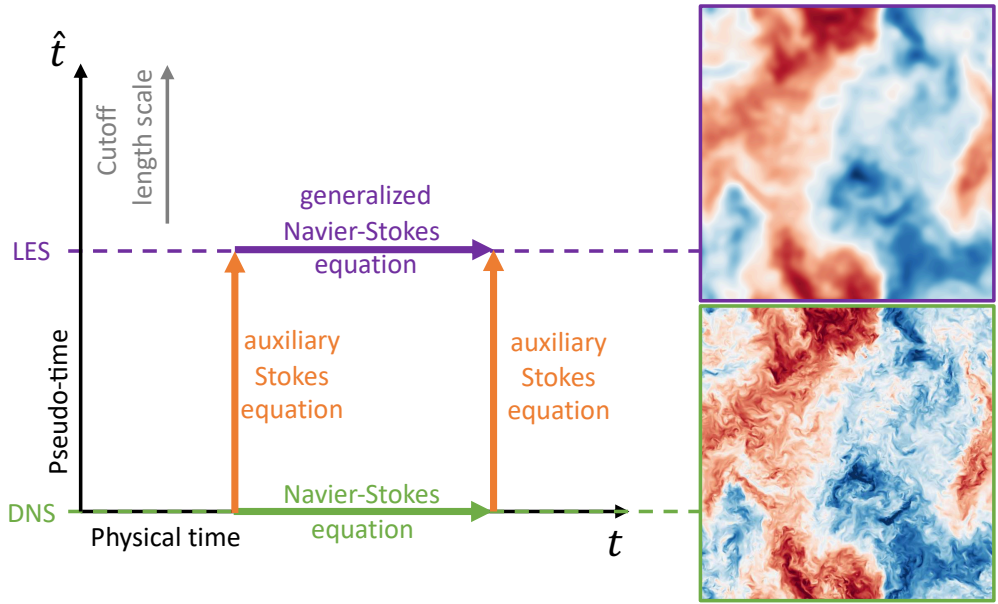}
	\caption{For physics-inspired coarsening for LES, a generalized velocity field is defined in as a function of physical time ($t$) and pseudo-time ($\hat{t} \sim \text{grid-size}^2$). The generalized Navier-Stokes equation for the LES velocity in physical time requires a closure model.}
	\label{fig:schematic}
\end{figure}

For the present purposes, an incompressible flow of a Newtonian fluid is considered. A velocity vector field, $\mathbf{u} = \mathbf{u}(\mathbf{x},t)$, evolves as a function of space, $\mathbf{x}$, and time, $t$, according to the Navier-Stokes equation,
\begin{equation}
	\frac{\partial u_i}{\partial t} + \frac{\partial}{\partial x_j}\left[ u_i u_j + p \delta_{ij} - \nu \left( \frac{\partial u_i}{\partial x_j} + \frac{\partial u_j}{\partial x_i} \right) \right] = 0,
	\hspace{0.1\linewidth}
	\frac{\partial u_j}{\partial x_j} = 0.
	\label{eq:Navier-Stokes}
\end{equation}
The pressure (divided by density), $p(\mathbf{x},t)$, enforces the divergence-free condition. A Poisson equation for the pressure follows from the divergence of the Navier-Stokes equation,
\begin{equation}
	\nabla^2 p = - \frac{\partial u_i}{\partial x_j} \frac{\partial u_j}{\partial x_i}.
	\label{eq:pressure-Poisson}
\end{equation}
Thus, a fully-resolved simulation (DNS) evolves Eq.\ \eqref{eq:Navier-Stokes} together with Eq.\ \eqref{eq:pressure-Poisson} in physical time with pseudo-time fixed at zero, $\hat{t} = 0$, indicated by the green arrow in Figure \ref{fig:schematic}.

\subsection{Auxiliary Stokes equation}
For the physics-inspired approach, a generalized velocity field, $\mathbf{w} = \mathbf{w}(\mathbf{x}, t; \hat{t})$, is defined as a function of both time, $t$, and a pseudo-time, $\hat{t}$.
The generalized velocity field at $\hat{t} = 0$ corresponds to the physical velocity field,
\begin{equation}
	\mathbf{w}(\mathbf{x},t; 0) = \mathbf{u}(\mathbf{x},t).
	\label{eq:pseudo-IC-velocity}
\end{equation}
As $\hat{t}$ increases, small-scale motions are removed from the generalized velocity field. Thus, the pseudo-time is an indicator of the resolution length scale of the generalized velocity.
To accomplish this, an auxiliary Stokes equation may be introduced to govern the pseudo-time evolution,
\begin{equation}
	\frac{\partial w_i}{\partial \hat{t}} + \frac{\partial}{\partial x_j}\left[ \hat{p}\delta_{ij} - \hat{\nu} \left( \frac{\partial w_i}{\partial x_j} + \frac{\partial w_j}{\partial x_i} \right) \right] = 0,
	\hspace{0.1\linewidth}
	\frac{\partial w_j}{\partial x_j} = 0.
	\label{eq:aux-Stokes}
\end{equation}
The nonlinear term of the Navier-Stokes equation ($u_i u_j$) is removed for Eq. \eqref{eq:aux-Stokes}, so there is no energy cascade in pseudo-time and hence no resupply of energy to small-scales dissipated by pseudo-viscosity, $\hat{\nu}(\mathbf{x})$.
The length scale at which the flow is smoothed may be heuristically estimated as $\ell \sim \sqrt{\hat{\nu} \hat{t}}$.
A pseudo-pressure, $\hat{p}(\mathbf{x},t;\hat{t})$, is also introduced to enforce a divergence-free condition for the generalized velocity at all pseudo-times. The divergence of the auxiliary Stokes equation is a Poisson equation for the pseudo-presssure,
\begin{equation}
	\nabla^2 \hat{p}  = \frac{\partial^2}{\partial x_i \partial x_j}\left[ \hat{\nu} \left( \frac{\partial w_i}{\partial x_j} + \frac{\partial w_j}{\partial x_i} \right) \right].
	\label{eq:pseudo-Poisson}
\end{equation}

Thus, a DNS may be artificially coarsened by, instead of applying a spatial filter, advancing Eq.\ \eqref{eq:aux-Stokes} in pseudo-time together with Eq.\ \eqref{eq:pseudo-Poisson} having initial conditions set by Eq. \eqref{eq:pseudo-IC-velocity}. This is represented in Figure \ref{fig:schematic} by the orange arrows.
In the case of a moving reference frame, the coordinate system only advances in physical time. It is frozen in pseudo-time by definition, so that Galilean invariance is satisfied.
Note that the pseudo-viscosity is distinct from the idea of an eddy viscosity (or turbulent viscosity) often used to model turbulent flows. The pseudo-viscosity acts only in pseudo-time and simply enables the coarsening procedure that is an alternative to spatial filtering. An eddy viscosity acts in physical time and attempts to approximate the evolution of a coarsened velocity field in physical time, which is considered next.
%It may not be necessary for compressible flows, see \S\ref{sec:compressible}.

\subsection{Generalized Navier-Stokes equation}
Explicit filtering is not typically used in practice for LES. Similarly, it is not envisioned at present that PIC-based LES practice will (necessarily) include the numerical solution of an auxiliary Stokes equation. The significance of Eq. \eqref{eq:aux-Stokes}, rather, is its implication for the evolution of the generalized velocity in physical time, keeping $\hat{t} \sim \ell^2$ fixed based on the grid resolution. This is illustrated by the purple arrow in Figure \ref{fig:schematic} and corresponds to establishing the effective equations for solving a coarsened representation of the flow, which may in practice be done on a coarser grid.

The outcome of the auxiliary Stokes equation ensures that the Navier-Stokes equation must be altered to accurately describe the evolution of the coarsened flow field in physical time. Thus, an effective evolution equation for LES may be derived by finding the appropriate residual force, $\mathbf{F}$, such that,
\begin{equation}
	\frac{\partial w_i}{\partial t} + \frac{\partial}{\partial x_j}\left[ w_i w_j + \phi\delta_{ij} - \nu \left( \frac{\partial w_i}{\partial x_j} + \frac{\partial w_j}{\partial x_i} \right) \right] = F_i,
	\hspace{0.1\linewidth}
	\frac{\partial w_j}{\partial x_j} = 0.
	%\label{eq:gen-NavierStokes}
\end{equation}
When $\hat{t} = 0$, the generalized velocity does satisfy Eq.\ \eqref{eq:Navier-Stokes} because of the initial condition, Eq.\ \eqref{eq:pseudo-IC-velocity}. This means that, $\mathbf{F}(\mathbf{x}, t; 0) = 0$.

Because the auxiliary Stokes equation globally conserves momentum in the absence of fluxes through domain boundaries, the residual force must likewise conserve momentum, and it may be written as the divergence of a residual stress tensor,
\begin{equation}
	F_ i = \frac{\partial \sigma_{ij}}{\partial x_j}.
\end{equation}
Thus, the evolution of the generalized velocity in physical time is given by the generalized Navier-Stokes equation,
\begin{equation}
	\frac{\partial w_i}{\partial t} + \frac{\partial}{\partial x_j}\left[ w_i w_j + \sigma_{ij} + \phi\delta_{ij} - \nu \left( \frac{\partial w_i}{\partial x_j} + \frac{\partial w_j}{\partial x_i} \right) \right] = 0,
	\hspace{0.1\linewidth}
	\frac{\partial w_j}{\partial x_j} = 0.
	\label{eq:gen-Navier-Stokes}
\end{equation}
Note that the divergence-free constraint on the auxiliary Stokes equation implies a divergence-free condition for the physical time evolution as well. This is enforced by a generalized pressure, $\phi(\mathbf{x},t;\hat{t})$, that is distinct from the pseudo-pressure, $\hat{p}$, which acts only during pseudo-time evolution.
The generalized pressure may be found by solution of the following Poisson equation,
\begin{equation}
	\nabla^2 \phi = - \frac{\partial w_i}{\partial x_j} \frac{\partial w_j}{\partial x_i} - \frac{\partial^2 \sigma_{ij}}{\partial x_i \partial x_j}.
	\label{eq:gen-Poisson}
\end{equation}

Equation \eqref{eq:gen-Navier-Stokes}, together with Eq. \eqref{eq:gen-Poisson} is unclosed in general. The residual stress tensor field must be approximated in terms of the generalized velocity field in order to solve for its evolution. More specifically, only the deviatoric component of $\boldsymbol{\sigma}$ is material to the closure problem, because any isotropic component may be lumped in with the generalized pressure and found via Eq. \eqref{eq:gen-Poisson}. For consistency, it is required that the generalized velocity, $\mathbf{w}$, evolve according to the Navier-Stokes equation for $\hat{t} = 0$. That is, Eq. \eqref{eq:gen-Navier-Stokes} must be identical to Eq. \eqref{eq:Navier-Stokes} when $\hat{t} = 0$. Thus,
\begin{equation}
	\phi(\mathbf{x}, t; 0) = p(\mathbf{x},t),
	\hspace{0.1\linewidth}
	\sigma_{ij}(\mathbf{x}, t; 0) = 0.
\end{equation}

\subsection{Residual stress tensor equation}
The significance of the above theory is its implications for the mathematical definition of the residual stress tensor, which is found by requiring consistency between the pseudo-time and physical time evolution of the generalized velocity field sketched in Figure \ref{fig:schematic}.
%The familiar form from spatial filtering theory, $\tau_{ij} = \overline{u_i u_j} - \overline{u}_i \overline{u}_j$, no longer applies.
%Instead, the auxiliary Stokes equation establishes the equation for $\tau_{ij}$ together with the generalized Navier-Stokes equation.
% TODO: make a new sketch?
Specifically, $\boldsymbol{\sigma}$ may be defined by asserting that advancing the generalized velocity field from $(t_0, \hat{t}_0)$ to $(t_f, \hat{t}_f)$ must be independent of the path in $t$-$\hat{t}$ space. Breaking the evolution in $t$-$\hat{t}$ space into infinitesimal increments, this requirement may be written as,
\begin{equation}
	\frac{\partial}{\partial \hat{t}}\left( \frac{\partial w_i}{\partial t} \right)
	=
	\frac{\partial}{\partial t}\left( \frac{\partial w_i}{\partial \hat{t}} \right),
	\label{eq:mixed-partial}
\end{equation}
which is simply the symmetry of mixed partial derivatives. More detailed treatment of the implications of this constraint for the form of the residual stress tensor will be given later.
% (Schwartz' theorem or Clairaut's theorem).

\subsection{Kinetic energy equations}

The scale-wise dynamics of kinetic energy is crucial to the representation of turbulent flows. The generalized kinetic energy (per unit mass) in PIC is given by $\tfrac{1}{2} w_i w_i$. First, consider how kinetic energy evolves in pseudo-time. Multiplying the auxiliary Stokes equation, Eq. \eqref{eq:aux-Stokes}, by $w_i$ and using a product rule on the viscous term,
\begin{equation}
	\frac{\partial (\tfrac{1}{2} w_i w_i)}{\partial \hat{t}} + \frac{\partial}{\partial x_j}\left[ \hat{p} w_j - 2 \hat{\nu} w_i S_{ij} \right] = - 2 \hat{\nu} S_{ij} S_{ij},
	\label{eq:aux-kinetic-energy}
\end{equation}
where $S_{ij} = \tfrac{1}{2}( \partial w_i/ \partial x_j + \partial w_j/\partial x_i)$ is the generalized strain-rate tensor. As pseudo-time increases, kinetic energy is removed by the pseudo-viscosity at a rate $2 \hat{\nu} S_{ij} S_{ij} \geq 0$.

The dynamics of generalized kinetic energy in physical time (at fixed, finite pseudo-time) can be written by multiplying the generalized Navier-Stokes equation, Eq.\ \eqref{eq:gen-Navier-Stokes} by $w_i$,
\begin{equation}
	\frac{\partial (\tfrac{1}{2} w_i w_i)}{\partial t} + \frac{\partial}{\partial x_j}\left[ \frac{1}{2} w_i w_i w_j + w_i \sigma_{ij} + \phi w_j - 2 \nu w_i S_{ij} \right] = \sigma_{ij} S_{ij}  - 2 \nu S_{ij} S_{ij}.
	\label{eq:gen-kinetic-energy}
\end{equation}
In addition to direct viscous dissipation, $2 \nu S_{ij} S_{ij} \geq 0$, the generalized kinetic energy is removed at a rate $\Pi = -\sigma_{ij} S_{ij}$. Note that $\Pi$ may be positive or negative in general, but that the expectation for turbulence dynamics is a preference toward $\Pi > 0$ representing the kinetic energy cascade. That is, $\Pi$ represents the rate at which kinetic energy associated with motions larger than $\ell \sim \sqrt{\hat{\nu} \hat{t}}$ is passed to motions smaller than $\ell$. Thus, an important consideration for any closure model for $\sigma_{ij}$ is its ability to remove energy at the proper rate to mimic the energy cascade.

\subsection{Juxtaposition with spatial filtering}

The traditional approach to LES theory is based on developing equations for the filtered velocity field \citep{Leonard1975, Germano1992},
\begin{equation}
	\overline{u}_i^\ell(\mathbf{x}, t) = \iiint G^\ell(\mathbf{r}) u_i(\mathbf{x} + \mathbf{r}, t) d\mathbf{r},
	\hspace{0.05\linewidth}
	\mathcal{F}\left\lbrace \overline{u}_i^\ell \right\rbrace  = 
	\left(2\pi\right)^3\mathcal{F}\left\lbrace G^\ell\right\rbrace
	\mathcal{F}\left\lbrace u_i \right\rbrace,
	\label{eq:spatial-filter}
\end{equation}
The relative width of the filter is $\ell$, with larger values of $\ell$ resulting in coarser fields. The notation $\mathcal{F}$ is used for a Fourier transform in three dimensions. The definition used here for an arbitrary field $a$ is,
\begin{equation}
	\mathcal{F}\left\lbrace a \right\rbrace(\mathbf{k}) = \frac{1}{\left(2\pi\right)^3} \iiint a(\mathbf{x}) \exp\left( -i \mathbf{k}\cdot\mathbf{x} \right) d\mathbf{x}.
	\label{eq:Fourier-transform}
\end{equation}
The three-dimensional inverse Fourier transform for an arbitrary function $b$ is therefore,
\begin{equation}
	\mathcal{F}^{-1}\left\lbrace b \right\rbrace(\mathbf{x}) = \iiint b(\mathbf{k}) \exp\left( i \mathbf{k}\cdot\mathbf{x} \right) d\mathbf{k}.
	\label{eq:inverse-Fourier-transform}
\end{equation}
This Fourier transform definition is commonly used for turbulence theory, e.g., \citep{Tennekes1972, Monin1975, Pope2000}.

LES practitioners rarely rely on a specific filter kernel shape. Common filter shapes used for theoretical appraisals include top-hat, spectral cutoff, and Gaussian kernels \citep{Borue1998}. For example, the Gaussian filter kernel is given by
\begin{equation}
	G^\ell(\mathbf{r}) = \frac{1}{(2\pi \ell)^{3/2}} \exp\left( - \frac{|\mathbf{r}|^2}{2\ell^2}\right),
	~~~~~~~
	\left(2\pi\right)^3\mathcal{F}\{G^\ell\}(\mathbf{k}) = \exp\left( -\frac{1}{2} |\mathbf{k}|^2 \ell^2 \right),
	\label{eq:Gaussian-kernel}
\end{equation}
The Gaussian filter provides a balanced trade-off between localization in physical space and wavenumber space, with exponential-of-square dropoff in both.

For a more general class of filter shapes, the Navier-Stokes equation, Eq.\ \eqref{eq:Navier-Stokes}, may be filtered to yield an equation for the filtered velocity field,
\begin{equation}
	\frac{\partial \overline{u}_i^\ell}{\partial t} + \frac{\partial}{\partial x_j}\left[ \overline{u}_i^\ell \overline{u}_j^\ell + \tau_{ij}^\ell + \overline{p}^\ell\delta_{ij} - \nu \left( \frac{\partial \overline{u}_i^\ell}{\partial x_j} + \frac{\partial \overline{u}_j^\ell}{\partial x_i} \right) \right] = 0,
	\hspace{0.1\linewidth}
	\frac{\partial \overline{u}_j^\ell}{\partial x_j} = 0.
	\label{eq:filtered-Navier-Stokes}
\end{equation}
Here, the subfilter stress tensor, $\tau_{ij}^\ell = \overline{u_i u_j}^\ell - \overline{u}_i^\ell \overline{u}_j^\ell$, plays a role analogous to that of the residual stress tensor in PIC, $\sigma_{ij}$, though their mathematical definitions may differ. Also, the filtered pressure, $\overline{p}^\ell$, is analogous to the generalized pressure in PIC, $\phi$.

The kinetic energy of the filtered velocity field evolves as,
\begin{equation}
	\frac{\partial (\tfrac{1}{2} \overline{u}_i^\ell \overline{u}_i^\ell)}{\partial t} + \frac{\partial}{\partial x_j}\left[ \frac{1}{2} \overline{u}_i^\ell \overline{u}_i^\ell \overline{u}_j^\ell  + \overline{u}_i \tau_{ij}^\ell  + \overline{p}^\ell \overline{u}_j^\ell - 2 \nu \overline{u}_i^\ell \overline{s}_{ij}^\ell \right] = \tau_{ij}^\ell \overline{s}_{ij}^\ell - 2 \nu \overline{s}_{ij}^\ell \overline{s}_{ij}^\ell.
	\label{eq:filtered-kinetic-energy}
\end{equation}
The strain-rate tensor is $s_{ij} = \tfrac{1}{2}\left( \partial u_i / \partial x_j + \partial u_j / \partial x_i \right)$ and the filtered strain-rate tensor is $\overline{s}_{ij}^\ell = \tfrac{1}{2} ( \partial \overline{u}_i^\ell / \partial x_j + \partial \overline{u}_j^\ell / \partial x_i )$. The two sinks of the filtered kinetic energy are analogous to those of the generalized kinetic energy in Eq.\ \eqref{eq:gen-kinetic-energy}.

Note that the preceding review of spatial filtering theory has implicitly assumed a spatially uniform filtering operation, i.e., $\ell \neq \ell(\mathbf{x})$. Specifically, the derivation of Eq.\ \eqref{eq:filtered-Navier-Stokes} by applying a filter operation to Eq.\ \eqref{eq:Navier-Stokes} relies on the commutative property of filtering with spatial differentiation, which does not hold if the filter width varies in space. In the case of spatially-varying filter width, additional commutator errors arise which (theoretically) require additional treatment \citep{Ghosal1995, Langford2001, Yalla2021, Moser2021}. The above derivation also presumed an unbounded domain in the filter definition. Section \ref{sec:wall-free} explores the similarity between PIC and filtering in the context of unbounded flows with uniform resolution. Even in this case, PIC provides useful insight. Then, section \ref{sec:complex} shows how PIC may be extended to more complex flows, including those which represent a challenge to the spatial filtering approach to LES theory.

\section{Unbounded flows with uniform resolution}
\label{sec:wall-free}

\subsection{Equivalence of PIC with Gaussian filtering}
For unbounded flows with uniform resolution (uniform $\hat{\nu}$), the pseudo-pressure satisfies a Laplace equation in free space, so $\hat{p} = 0$.
That is, the pseudo-pressure is not needed to enforce the divergence-free condition on the generalized velocity. In this case, the %generalized Navier-Stokes equation and 
auxiliary Stokes equation simplifies to,
%\begin{equation}
%	\frac{\partial w_i}{\partial t} = - w_j \frac{\partial w_i}{\partial x_j} - \frac{\partial \tau_{ij}}{\partial x_j} - \frac{\partial \overline{p}}{\partial x_i} + \nu \nabla^2 \overline{u}_i
%	\label{eq:gen-Navier-Stokes-free}
%\end{equation}
\begin{equation}
	\frac{\partial w_i}{\partial \hat{t}} = \hat{\nu} \nabla^2 w_i,
	\hspace{0.1\linewidth}
	w_i(\mathbf{x}, t; 0) =u_i^\ell(\mathbf{x}, t).
	\label{eq:aux-Stokes-free}
\end{equation}
The formal solution to Eq. \eqref{eq:aux-Stokes-free} in an unbounded domain is readily obtained
\begin{equation}
	w_i(\mathbf{x}, t; \hat{t})
	= \iiint \frac{1}{\left( 4 \pi \hat{\nu} \hat{t} \right)^{3/2}} \exp\left( - \frac{|\mathbf{r}|^2}{4 \hat{\nu} \hat{t}} \right) u_i(\mathbf{x}+\mathbf{r}, t) d\mathbf{r}
	\equiv \overline{u}_i^\ell(\mathbf{x}, t),
\end{equation}
which is precisely a Gaussian filter with width $\ell = \sqrt{2\hat{\nu}\hat{t}}$. Therefore, the physics-inspired coarsening approach is precisely equivalent to spatial filtering with a Gaussian kernel for unbounded flows with uniform resolution. All the strengths and theoretical insights of spatial filtering naturally carry over, but a new perspective on the residual stress tensor also emerges.

% TODO: this goes later in the Gaussian filtering section
%Above wavenumbers $\kappa \sim \eta^{-1}$, the spectrum falls off as a stretched exponential (CITE). Thus, it is sufficient for direct numerical simulations (DNS) to have a grid spacing is set to $\Delta x \sim \eta$. For wavenumbers above $\kappa \sim \ell^{-1}$, the resulting spectrum falls off exponentially (or faster).

\subsection{PIC expression for the residual stress tensor}

With the equivalence to Gaussian filtering established for PIC of unbounded flows with uniform resolution, the mathematical definition of the residual stress tensor could be straightfowardly written as $\sigma_{ij} = \tau_{ij} \equiv \overline{u_i u_j}^\ell - \overline{u}_i^\ell \overline{u}_j^\ell$. However, insights beyond those arrived at via spatial filtering may be obtained by following the PIC logic further in the form of Eq.\ \eqref{eq:mixed-partial}.

In the following, flow subjected to an arbitrary (divergence-free) forcing function, $\mathbf{f}(\mathbf{x},t)$ is considered, so Eq. \eqref{eq:Navier-Stokes} becomes,
\begin{equation}
	\frac{\partial u_i}{\partial t} + \frac{\partial}{\partial x_j}\left[ u_i u_j + p \delta_{ij} - \nu \left( \frac{\partial u_i}{\partial x_j} + \frac{\partial u_j}{\partial x_i} \right) \right] = f_i,
	\hspace{0.1\linewidth}
	\frac{\partial u_j}{\partial x_j} = 0.
	\label{eq:forced-Navier-Stokes}
\end{equation}
The generalized Navier-Stokes equations are likewise supplemented with a generalized force, $\mathbf{g}(\mathbf{x},t)$, that represents the impact of the physical forcing on the coarsened flow representation,
\begin{equation}
	\frac{\partial w_i}{\partial t} + \frac{\partial}{\partial x_j}\left[ w_i w_j + \sigma_{ij} + \phi\delta_{ij} - \nu \left( \frac{\partial w_i}{\partial x_j} + \frac{\partial w_j}{\partial x_i} \right) \right] = g_i,
	\hspace{0.1\linewidth}
	\frac{\partial w_j}{\partial x_j} = 0.
	\label{eq:gen-forced-Navier-Stokes}
\end{equation}

Substituting Eqs. \eqref{eq:aux-Stokes-free} and \eqref{eq:gen-forced-Navier-Stokes} into Eq. \eqref{eq:mixed-partial},
\begin{equation}
	\frac{\partial}{\partial \hat{t}}\left( \nu \nabla^2 w_i - \frac{\partial \phi}{\partial x_i} - \frac{\partial w_i w_j}{\partial x_j} - \frac{\partial \sigma_{ij}}{\partial x_j} + g_i \right)
	- \frac{\partial}{\partial t}\left( \hat{\nu} \nabla^2 w_i \right)
	= 0
\end{equation}
and rearranging in terms of time and pseudo-time derivatives,
\begin{equation}
	\nu \nabla^2 \left(\frac{\partial w_i}{\partial \hat{t}}\right)
	- \frac{\partial}{\partial x_i}\left(\frac{\partial \phi}{\partial \hat{t}}\right)
	- \frac{\partial}{\partial x_j}\left(w_i  \frac{\partial w_j}{\partial \hat{t}} + w_j \frac{\partial w_i}{\partial \hat{t}}\right)
	- \frac{\partial}{\partial x_j}\left( \frac{\partial \sigma_{ij}}{\partial \hat{t}} \right)
	+ \frac{\partial g_i}{\partial \hat{t}}
	= \hat{\nu} \nabla^2 \left(\frac{\partial w_i}{\partial t} \right)
\end{equation}
and further substitution of Eqs. \eqref{eq:aux-Stokes-free} and \eqref{eq:gen-forced-Navier-Stokes},
\begin{multline}
	\nu \nabla^2 \left(\hat{\nu} \nabla^2 w_i \right)
	- \frac{\partial}{\partial x_i}\left(\frac{\partial \phi}{\partial \hat{t}}\right)
	- \frac{\partial}{\partial x_j}\left(\hat{\nu} w_i \nabla^2 w_j + \hat{\nu} w_j \nabla^2 w_i \right)
	- \frac{\partial}{\partial x_j}\left( \frac{\partial \sigma_{ij}}{\partial \hat{t}} \right)
	+ \frac{\partial g_i}{\partial \hat{t}} \\
	= \hat{\nu} \nabla^2 \left( \nu \nabla^2 w_i - \frac{\partial \phi}{\partial x_i} - \frac{\partial w_i w_j}{\partial x_j} - \frac{\partial \sigma_{ij}}{\partial x_j} + g_i \right).
\end{multline}
Now, further simplification leads to,
\begin{equation}
	- \frac{\partial}{\partial x_j}\left( \frac{\partial \sigma_{ij}}{\partial \hat{t}} - \hat{\nu} \nabla^2 \sigma_{ij} - 2 \hat{\nu} \frac{\partial w_i}{\partial x_k} \frac{\partial w_j}{\partial x_k} + \frac{\partial \phi}{\partial \hat{t}} \delta_{ij} - \hat{\nu}\nabla^2 \phi \delta_{ij} \right) = \frac{\partial g_i}{\partial \hat{t}} - \hat{\nu} \nabla^2 g_i.
	\label{eq:forced-constraint}
\end{equation}
Equation \eqref{eq:forced-constraint} simply expresses the condition necessary for Eq.\ \eqref{eq:gen-forced-Navier-Stokes} to correctly describe the large-scale dynamics embodied in the generalized velocity at finite pseudo-time. Because the residual stress should be disentangled from the (arbitrary) forcing function, both sides for Eq.\ \eqref{eq:forced-constraint} should be set to zero. Setting the right-hand side to zero, the generalized forcing function may be recognized as the Gaussian-filtered force, c.f. Eq.\ \eqref{eq:aux-Stokes-free},
\begin{equation}
	\frac{\partial g_i}{\partial \hat{t}} = \hat{\nu} \nabla^2 g_i,
	\hspace{0.1\linewidth}
	g_i(\mathbf{x}, t; 0) = f_i(\mathbf{x}, t).
\end{equation}
The left-hand side of Eq. \eqref{eq:forced-constraint}, once set to zero, may be simplified. Namely, the terms involving the generalized pressure, $\phi$, may be removed on either of two considerations. First, the generalized pressure may be identified with the Gaussian-filtered pressure,
\begin{equation}
	\frac{\partial \phi}{\partial \hat{t}} = \hat{\nu} \nabla^2 \phi,
	\hspace{0.1\linewidth}
	\phi(\mathbf{x}, t; 0) = p(\mathbf{x}, t).
\end{equation}
Alternatively, and more generally, the dynamics of the generalized velocity depend only on the deviatoric part of the residual stress tensor, so any isotropic contribution can be safely ignored.

%Any isotropic contribution to $\sigma_{ij}$ may be included with the pressure in the Poisson equation.
%Thus, allowing that addition of a divergence-free tensor does not affect Eq. \eqref{eq:gen-Navier-Stokes},
%Without loss of generality, it may be taken that
%\begin{equation}
%	\frac{\partial \tau_{ij}}{\partial \hat{t}} - \hat{\nu} \nabla^2 \tau_{ij} - 2 \hat{\nu} \frac{\partial \overline{u}_i}{\partial x_k} \frac{\partial \overline{u}_j}{\partial x_k} + \frac{\partial \overline{p}}{\partial \hat{t}} \delta_{ij} - \hat{\nu}\nabla^2 \overline{p} \delta_{ij} = 0
%\end{equation}
%Further, any isotropic components of the residual stress tensor do not alter the evolution of the generalized velocity in physical time, because they may be merged with the generalized pressure and participate in the Poisson equation to enforce the divergence-free condition. As a result,
%only the deviatoric component, $\tau_{ij}^{(d)}$ alters the velocity field evolution, so we take the deviatoric part of the above equation,

A sufficient condition for satisfying Eq.\ \eqref{eq:forced-constraint} and thus Eq.\ \eqref{eq:mixed-partial} is
\begin{equation}
	\frac{\partial \sigma_{ij}}{\partial \hat{t}} = \hat{\nu} \nabla^2 \sigma_{ij} + 2 \hat{\nu} \frac{\partial w_i}{\partial x_k} \frac{\partial w_j}{\partial x_k}.
	\label{eq:residual-stress-diffusion}
\end{equation}
This is a forced diffusion equation in pseudo-time for $\boldsymbol{\sigma}$, and its formal solution in unbounded space is also readily obtained 
\begin{equation}
	\sigma_{ij}(\mathbf{x}, t; \hat{t}) = \int_{0}^{\hat{t}} \left[ \iiint \frac{1}{\left( 4 \pi \hat{\nu} (\hat{t}-t^\prime) \right)^{3/2}} \exp\left( - \frac{|\mathbf{r}|^2}{4 \hat{\nu} (\hat{t} - t^\prime)} \right)
	\frac{\partial w_i}{\partial x_k}%(\mathbf{x}+\mathbf{r}, t; t^\prime)
	\frac{\partial w_j}{\partial x_k}(\mathbf{x}+\mathbf{r}, t; t^\prime)
	d\mathbf{r} \right] dt^\prime.
	\label{eq:residual-stress-formal}
\end{equation}
The residual stress at pseudo-time $\hat{t}$ may thus be interpreted as an integral over all earlier pseudo-times, $0 \leq t^\prime \leq t$, of the velocity gradient product at $t^\prime$ smoothed by a pseudo-viscosity from its earlier pseudo-time up until the pseudo-time at which the residual stress is evaluated, i.e., over the pseudo-time range $\hat{t} - t^\prime$.

%and interpreted in terms of multiscale velocity gradient products, see \citet{Johnson2021b} for details.

%That is, the residual stress tensor is a superposition of Gaussian filtered generalized velocity gradient products from $0 \leq t^\prime \leq \hat{t}$. The interpretation of this result is carried out in detail by \citet{Johnson2021b}. In short, the residual stress tensor is written in terms of velocity gradients at all smaller scales, $0 \leq 2 \hat{\nu} t^\prime \leq 2 \hat{\nu} \hat{t}$.

%\begin{itemize}
%	\item TODO: discuss energetics?
%\end{itemize}

\subsection{Insights into the energy cascade}

The equivalence of PIC with Gaussian spatial filtering may be invoked, in the case of unbounded flows with uniform resolution, to rephrase the subfilter stress tensor of filtering theory in terms of multiscale velocity gradients. Dividing Eq.\ \eqref{eq:residual-stress-diffusion} by $2 \hat{\nu}$, with the established relation $\ell^2 = 2 \hat{\nu} \hat{t}$,
\begin{equation}
	\frac{\partial \tau_{ij}}{\partial (\ell^2)} = \frac{1}{2} \nabla^2 \tau_{ij} + \frac{\partial \overline{u}_i^\ell}{\partial x_k} \frac{\partial \overline{u}_j^\ell}{\partial x_k}
\end{equation}
therefore has a formal solution, Eq.\ \eqref{eq:residual-stress-formal}, that can itself be written in terms of an integral of Gaussian filtered velocity gradients at all scales $0 \leq \sqrt{\alpha} \leq \ell$,
\begin{equation}
	\tau_{ij} = \int_{0}^{\ell^2} \overline{\frac{\partial \overline{u}_i^{\sqrt{\alpha}}}{\partial x_k} \frac{\partial \overline{u}_j^{\sqrt{\alpha}}}{\partial x_k}}^{\beta} d\alpha
\end{equation}
where $\beta = \sqrt{\ell^2 - \alpha}$ is the width of the complementary filter which smooths from scale $\sqrt{\alpha}$ to scale $\ell$.

Invoking the definition of the generalized second moment from \citet{Germano1992},
\begin{equation}
	\tau^\ell(a,b) = \overline{a b}^\ell - \overline{a}^\ell \overline{b}^\ell,
\end{equation}
of which the subfilter stress tensor is one special case $\tau_{ij} = \tau(u_i, u_j)$, the PIC-based phrasing of the Gaussian filter's stress tensor can be split into resolved scale and subfilter scale contributions,
\begin{equation}
	\tau_{ij} = \ell^2 \frac{\partial \overline{u}_i^\ell}{\partial x_k} \frac{\partial \overline{u}_j^\ell}{\partial x_k} + \int_{0}^{\ell^2} \tau^\beta\left(  \frac{\partial \overline{u}_i^{\sqrt{\alpha}}}{\partial x_k}, \frac{\partial \overline{u}_j^{\sqrt{\alpha}}}{\partial x_k} \right)d\alpha.
	\label{eq:subfilter-stress-formal}
\end{equation}
The first of the two terms on the right hand side is the nonlinear gradient model, which by itself performs well in \textit{a priori} testing compared with eddy viscosity models \citep{Clark1979, Borue1998}. The second term shows how such a model leaves out smaller-scale (unresolved) content. Further decomposing the filtered velocity gradients in Eq.\ \eqref{eq:subfilter-stress-formal} into strain-rate and vorticity, and forming the product $\Pi^\ell = -\tau_{ij}^\ell \overline{s}_{ij}^\ell$, the local energy cascade rate may be written in terms of vorticity stretching and strain self-amplification,
\begin{equation}
	\Pi^\ell = \Pi_{s1}^\ell + \Pi_{\omega 1}^\ell + \Pi_{s2}^\ell + \Pi_{\omega 2}^\ell + \Pi_{c}^\ell,
	\label{eq:Pi-decomposition}
\end{equation}
where
\begin{equation}
	\Pi_{s1}^\ell
	= -\ell^2 \overline{s}_{ij}^\ell  \overline{s}_{jk}^\ell  \overline{s}_{ki}^\ell
	= \text{strain-rate self-amplification at scale~} \ell,
	\label{eq:resolved-strain-amp}
\end{equation}
\begin{equation}
	\Pi_{\omega 1}^\ell
	= \frac{1}{4} \ell^2 \overline{s}_{ij}^\ell \overline{\omega}_{i}^\ell \overline{\omega}_j^\ell
	= \text{vorticity stretching at scale~} \ell,
	\label{eq:resolved-vorticity-stretch}
\end{equation}
\begin{equation}
	\Pi_{s2}^\ell = - \overline{s}_{ij}^\ell \int_{0}^{\ell^2}  \tau_{\beta}\left( \overline{s}_{jk}^{\sqrt{\alpha}}, \overline{s}_{ki}^{\sqrt{\alpha}} \right) ~ d\alpha
	%= - \int_{0}^{\ell^2} d\alpha~ \overline{S}_{ij}^\ell \tau_{\beta}\left( \overline{S}_{jk}^{\sqrt{\alpha}}, \overline{S}_{ki}^{\sqrt{\alpha}} \right).
	= \text{multiscale strain amplification by strain at ~} \ell,
	\label{eq:multiscale-strain-amp}
\end{equation}
\begin{equation}
	\Pi_{\omega 2}^\ell
	= \overline{s}_{ij}^\ell \int_{0}^{\ell^2} \tau_\beta\left(\overline{\omega}_i^{\sqrt{\alpha}}, \overline{\omega}_j^{\sqrt{\alpha}}\right) ~ d\alpha
	%= \int_{0}^{\ell^2} d\alpha~\overline{S}_{ij}^\ell \tau_\beta\left(\overline{\omega}_i^{\sqrt{\alpha}}, \overline{\omega}_j^{\sqrt{\alpha}}\right).
	= \text{multiscale vorticity stretching by strain at~} \ell,
	\label{eq:multiscale-vorticity-stretch}
\end{equation}
\begin{equation}
	\Pi_{c}^\ell = 2 \overline{s}_{ij}^\ell \int_{0}^{\ell^2}  \tau_\beta\left( \overline{s}_{jk}^{\sqrt{\alpha}}, \overline{\Omega}_{ki}^{\sqrt{\alpha}} \right) ~ d\alpha
	= \text{multiscale vortex thinning by strain at~} \ell.
	\label{eq:multiscale-cross}
\end{equation}
In this way, the rephrasing of spatial filtering in terms of PIC for unbounded flows with uniform resolution leads to a unique theoretical insight: the exact relation of commonly invoked mechanisms to the energy cascade. For more details, the reader is referred to \citet{Johnson2020, Johnson2021b}. A treatment of the topic for a broader audience is given in \citet{Johnson2021a}.

% TODO: talk about 50-50 split and eddy viscosity like nature of the multiscale terms

%Ultimately, for unbounded incompressible turbulent flows with homogeneous isotropic spatial resolution, this viscosity-inspired coarsening method is simply a rephrasing of spatial filtering with the specific choice of a Gaussian filter kernel. Nevertheless, this rephrasing enables significant physical insight: an exact spatially-local relationship between vortex stretching, strain self-amplification and the kinetic energy cascade . %To the authors knowledge, no other analysis method provides such an insight.

Direct numerical simulations of homogeneous isotropic turbulence demonstrate that the multiscale gradient terms in $\boldsymbol{\tau}$ align closely with the filtered strain-rate tensor \citep{Johnson2021b}. The eddy viscosity model is thus a good approximation for the second term on the right hand side of Eq.\ \eqref{eq:subfilter-stress-formal}. Overall, this suggests the applicability of a mixed model, with $\boldsymbol{\sigma} = \boldsymbol{\tau}$ being the sum of nonlinear gradient and eddy viscosity terms \citep{Vreman1994a}.

\subsection{An alternative to the Germano identity}
The dynamic procedure \citep{Germano1991} based on the Germano identity \citep{Germano1992} is one of the most common direct uses of spatial filtering theory in LES practice. The basic idea is that coefficients for a given model may be estimated from the resolved scales using a test filter larger than the grid size. A similar procedure may be developed with physics-inspired coarsening via the dependence of $\boldsymbol{\sigma}$ on the cutoff scale (i.e., pseudo-time).

This may be briefly demonstrated for a generic eddy viscosity model,
\begin{equation}
	\sigma_{ij}^{(d)} \approx - 2 \nu_T S_{ij}.
	\label{eq:eddy-viscosity}
\end{equation}
First, Eq. \eqref{eq:eddy-viscosity} is substituted into Eq. \eqref{eq:residual-stress-diffusion} and the spatial variability of $\nu_T$ is neglected (for convenience, as commonly done for the Smagorinsky coefficient),
\begin{equation}
	-2 S_{ij} \frac{\partial \nu_T}{\partial \hat{t}} = 2 \hat{\nu} \left( \frac{\partial w_i}{\partial x_k} \frac{\partial w_j}{\partial x_k} - \frac{1}{3} \frac{\partial w_m}{\partial x_n} \frac{\partial w_m}{\partial x_n} \delta_{ij} \right).
\end{equation}
%Note that the spatial variability of the Smagorinski coefficient is typically neglected in the most common dynamic model implementations.
The above expression may be modified to the Smagorinski form and the spatial variability of the filtered strain-rate may also be factored in, if desired.

Second, \cite{Kolmogorov1941a} scaling is assumed, $\nu_T \sim \ell^{4/3} \sim \hat{t}^{2/3}$, so that the variation of the eddy viscosity in pseudo-time may be evaluated, $\frac{\partial \nu_T}{\partial \hat{t}} = \frac{2}{3} \frac{\nu_T}{\hat{t}}$.
Then Eq. \eqref{eq:dyn-nut} becomes,
\begin{equation}
	-\frac{4}{3} S_{ij} \nu_T = \ell^2 \left( \frac{\partial w_i}{\partial x_k} \frac{\partial w_j}{\partial x_k} - \frac{1}{3} \frac{\partial w_m}{\partial x_n} \frac{\partial w_m}{\partial x_n} \delta_{ij} \right)
\end{equation}
where $\ell^2 = 2 \hat{\nu} \hat{t}$. This equation is over-determined because, in general, a scalar $\nu_T$ will not be found that can satisfy the full tensor equation.
Therefore, as is typical for dynamic procedures, a least-squares procedure is employed \citep{Lilly1992}.
This step leads to a projection on $S_{ij}$ that carries the physical meaning of matching the energy cascade rate.
The resulting expression for the eddy viscosity is,
\begin{equation}
	\nu_T
	= \frac{3 \ell^2}{4} \frac{A_{ik} A_{jk} A_{ij}}{S_{mn} S_{mn}}
	= \frac{3\ell^2}{4} \frac{\frac{1}{4} W_i S_{ij} W_j - S_{ij} S_{jk} S_{ki}}{S_{mn} S_{mn}},
	\label{eq:dyn-nut}
\end{equation}
where $A_{ij} = \partial w_i / \partial x_j$ is the generalized velocity gradient tensor, $S_{ij}  = \tfrac{1}{2} (A_{ij} + A_{ji})$ is the generalized strain-rate tensor, and $W_{i} = \epsilon_{ijk} A_{kj}$ is the generalized vorticity.
The denominator of this relation is positive-definite. The numerator represents the sum of vortex stretching and strain-rate self-amplification, which is positive on average for turbulent flows. Local negative values of the numerator may necessitate averaging or clipping strategies as done in the Germano-based dynamic procedure. %Additional equations, such as the second derivative of $\tau_{ij}$ with $\hat{t}$ may be considered to avoid the assumption of Kolmogorov scaling, analogous to the use of a second test filter in the scale-dependent dynamic procedure (CITE).

\cite{Johnson2021b} demonstrated that the eddy viscosity approximation is more physically accurate for the only part of the residual stress tensor,
\begin{equation}
	\epsilon_{ij} = \sigma_{ij} - 2 \hat{\nu} \hat{t} \frac{\partial w_i}{\partial x_k} \frac{\partial w_j}{\partial x_k},
\end{equation}
which represents multiscale vortex stretching and strain-rate amplification in Eq. \eqref{eq:subfilter-stress-formal}. It may be shown from Eqs. \eqref{eq:aux-Stokes-free} and \eqref{eq:residual-stress-diffusion} that,
\begin{equation}
	\frac{\partial \epsilon_{ij}}{\partial \hat{t}} = \hat{\nu} \nabla^2 \epsilon_{ij} + 4 \hat{\nu}^2 \hat{t} \frac{\partial^2 w_i}{\partial x_m \partial x_n} \frac{\partial^2 w_j}{\partial x_m \partial x_n} 
\end{equation}
Substituting the eddy viscosity approximation, Eq.\ \eqref{eq:eddy-viscosity}, for $\epsilon_{ij}^{(d)}$ rather than the full residual stress tensor, $\sigma_{ij}^{(d)}$, and again assuming $\nu_T \sim \hat{t}^{2/3}$ with a least squares approach leads to a `dynamic' mixed model,
\begin{equation}
	\sigma_{ij}^{(d)} = \ell^2 \left( A_{ik} A_{jk} \right)^{(d)} - 2 \nu_T S_{ij},
	\hspace{0.1\linewidth}
	\nu_T = - \frac{3 \ell^4}{4} \frac{B_{ikl} S_{ij} B_{jkl}}{S_{mn} S_{mn}},
	\label{eq:dyn-mixed-model}
\end{equation}
where $B_{ijk} = \partial^2 w_i / \partial x_j \partial x_k$ and $\ell^2 = 2 \hat{\nu}\hat{t}$. The appearance of the second derivative of the generalized velocity field in the mixed model is less practical when numerical errors are inherent in assessing spatial derivatives. Nonetheless, Eq.\ \eqref{eq:dyn-mixed-model} provides a useful initial test for assessing the relative strengths of PIC theory.

\subsection{\textit{A priori} testing}

Direct numerical simulations of homogeneous isotropic turbulence are used for \textit{a priori} tests reported in this subsection.
The incompressible Navier-Stokes equation, Eq. \eqref{eq:Navier-Stokes}, is solved using a pseudo-spectral method with $1024$ collocation points in each direction. The fully-resolved velocity field, $\mathbf{u}$, is advanced in time with a second-order Adams-Bashforth scheme, and the pressure $p$ simply enforces the divergence-free condition. The $2\sqrt{2}/3$ rule for wave number truncation is used with phase-shift dealiasing \citep{Patterson1971}. The forcing, $\mathbf{f}$, is specifically designed to maintain constant kinetic energy in the first two wave number shells. 

\begin{figure}
	\centering
	\includegraphics[width=0.485\linewidth]{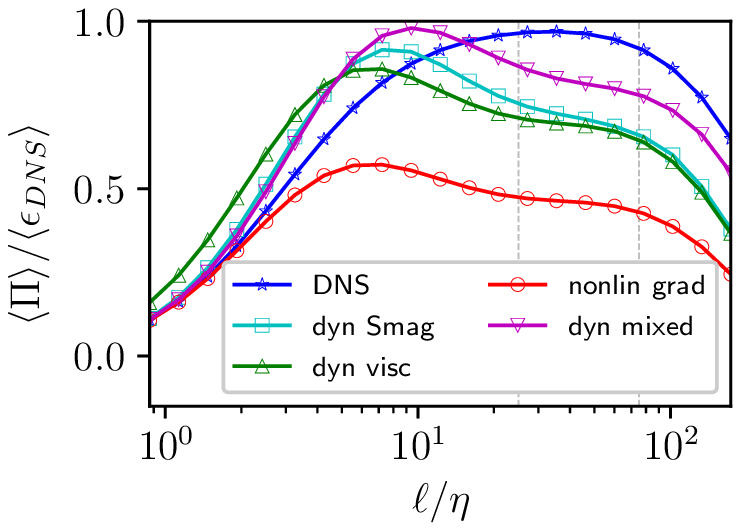}
	\includegraphics[width=0.505\linewidth]{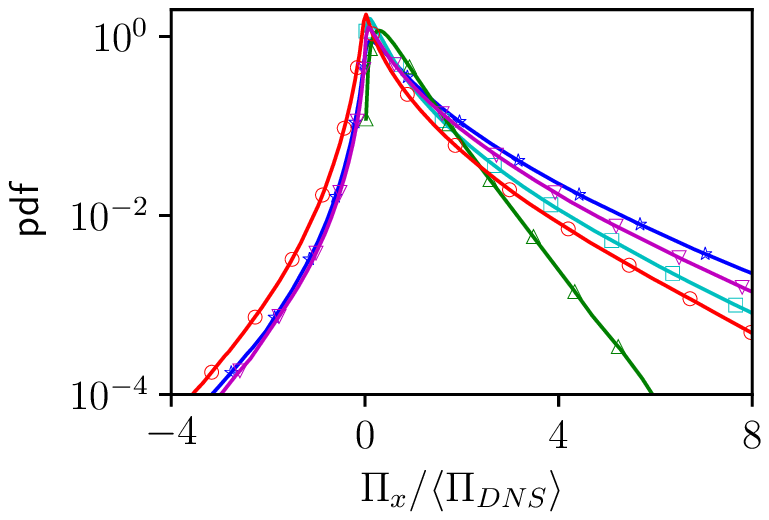}
	\vskip -2mm
	(a) \hspace{0.46\linewidth} (b)\\
	\includegraphics[trim = 0mm 8mm 0mm 0mm, clip, width=0.485\linewidth]{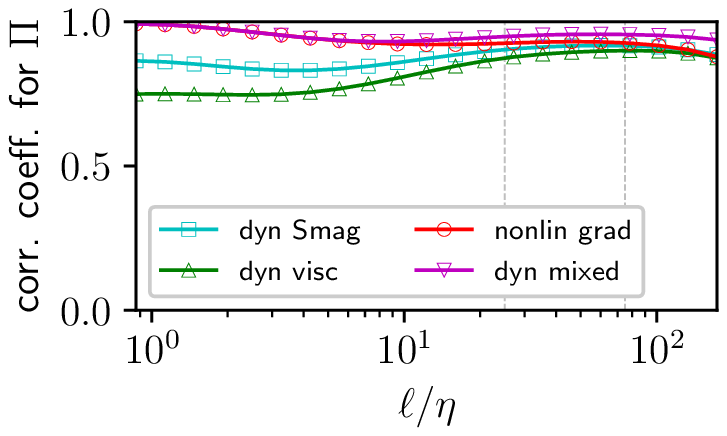}
	\includegraphics[trim = 0mm 8mm 0mm 0mm, clip, width=0.485\linewidth]{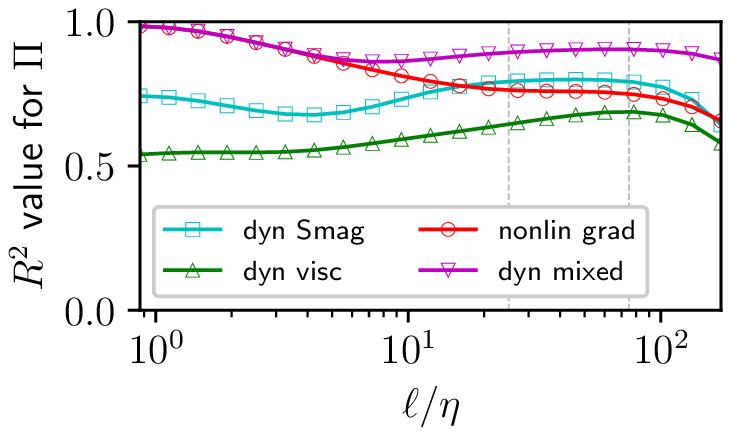}
	(c) \hspace{0.46\linewidth} (d)\\
	\includegraphics[width=0.485\linewidth]{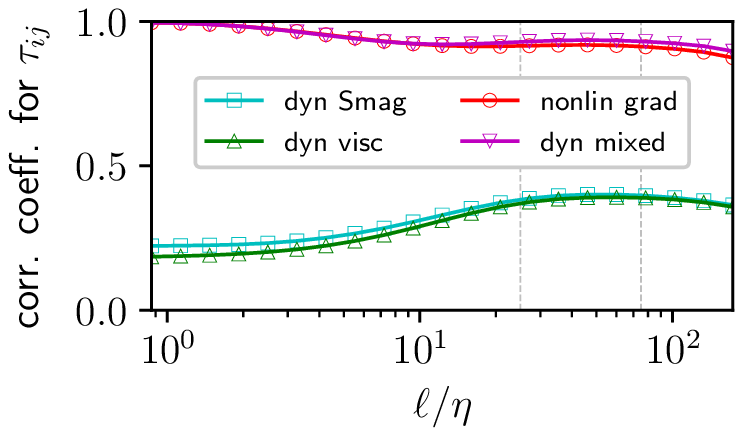}
	\includegraphics[width=0.485\linewidth]{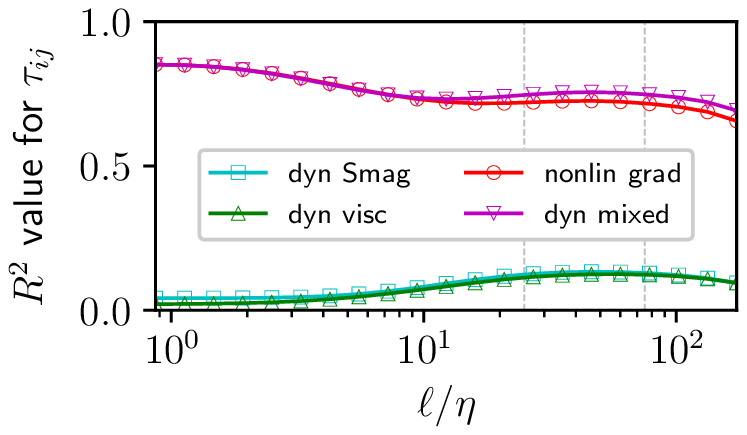}
	(e) \hspace{0.46\linewidth} (f)\\
	\caption{A priori comparison of new dynamic eddy viscosity (`dyn visc') and dynamic mixed (`dyn mixed') models with DNS results and existing models: dynamic Smagorinsky (`dyn Smag') and nonlinear gradient model (`nonlin grad'): (a) Mean cascade rate, (b) PDF of energy cascade rate at $\ell = 46 \eta$, (c-d) correlation coefficient and R$^2$ coefficient for the energy cascade rate, respectively, (e-f) correlation coefficient and R$^2$ coefficient for the full residual stress tensor, respectively. Two vertical dashed gray lines indicate rough bounds for the approximate inertial subrange of scales in the DNS, $25 \leq \ell/\eta \leq 75$.}
	\label{fig:germano_alternative}
\end{figure}

The simulation was initialized using a Gaussian random velocity field satisfying a model turbulent energy spectrum \citep{Pope2000}. The simulation was first run through a startup period to reach statistical stationarity. Then, statistics are computed over $6$ large-eddy turnover times. The Taylor-scale Reynolds number is approximately $Re_\lambda = 400$ with grid resolution $k_{\max} \eta = 1.4$. The integral length scale is about $20\%$ of the periodic box size of $2\pi$ and $L/\eta= 460$. The skewness of the longitudinal velocity gradient is $-0.58$, and the flatness of the longitudinal and transverse velocity gradients are $8.0$ and $12.4$, respectively, in reasonable agreement with previous simulations \citep{Ishihara2007}.
The \textit{a priori} assessment is carried out for the `dynamic' viscosity model, Eq.\ \eqref{eq:dyn-nut} and `dynamic' mixed model, Eq.\ \eqref{eq:dyn-mixed-model}, along with existing popular models including the standard dynamic Smagorinsky model \citep{Germano1991, Lilly1992} and nonlinear gradient model \citep{Clark1979, Borue1998}. For all three dynamic models, spatial averaging is used for the numerator and denominator of the Smagorinsky coefficient or eddy viscosity. The results are shown in Figure \ref{fig:germano_alternative}. For this case, PIC and Gaussian filtering are equivalent so the language of each approach is used interchangeably.

Perhaps the most important practical consideration for LES models is providing the proper rate of energy removal to unresolved scales. Figure \ref{fig:germano_alternative}a shows the the average cascade rate, $\Pi$, as a function of filter width, $\ell = \sqrt{2 \hat{\nu} \hat{t}}$, normalized by the average dissipation rate in the unfiltered DNS. The blue curve with star symbols shows that the cascade rate is nearly equal to the dissipation rate for $25 \lesssim \ell / \eta \lesssim 75$, an approximate inertial range for the simulation. The cascade rate decays to zero as $\ell$ decreases because viscous dissipation becomes significant near the Kolmogorov scale. The other four curves indicate the extent to which the four models reproduce the correct cascade rate when applied directly to coarsened DNS data.
As is well-known, the nonlinear gradient model is under-dissipative, especially in the inertial range, and the dynamic Smagorinsky model provides a better approximation of the energy removal rate. The PIC-based dynamic eddy viscosity model performs very similarly to the filter-based dynamic Smagorinsky model. The PIC-based dynamic mixed model is confirmed to remove energy at a higher rate in the inertial range than the nonlinear gradient or eddy viscosity models on their own.

The probability density function (PDF) of the energy cascade rate, $\Pi$, for each model at $\ell = 46 \eta$  is shown in Figure \ref{fig:germano_alternative}b along with the coarsened DNS result. The DNS data indicate the existence of some backscatter, but the distribution is strongly skewed toward positive cascade rates. The two eddy viscosity models do not allow any backscatter by construction due to the spatial averaging used as part of the dynamic procedure. The nonlinear gradient and PIC-based dynamic mixed model provide relatively appropriate probabilities of backscatter, with the mixed model having the best overall agreement with DNS.

Figures \ref{fig:germano_alternative}c and \ref{fig:germano_alternative}d show the correlation coefficient and R-squared value based on each models point-by-point prediction of the cascade rate, $\Pi$. An R-squared value of one indicates a perfect model. All four models are highly correlated with DNS in the inertial range, but the absolute error measured by the R-squared value shows more variation in the models performances. Both metrics show that the PIC-based dynamic mixed model is in closest agreement with DNS. Figures \ref{fig:germano_alternative}e and \ref{fig:germano_alternative}f show the correlation coefficients and R-squared values based on the full residual (subfilter) stress tensor. The stress tensor is known to align poorly with the strain-rate tensor, so the eddy viscosity models perform poorly for both metrics. The nonlinear gradient model performs quite well and the dynamic mixed model shows only slight improvement. The major benefit of the mixed model over the nonlinear gradient model has already been demonstrated in Figure \ref{fig:germano_alternative}a.

%The new dynamic viscosity procedure introduced here is comparable in accuracy to the original dynamic Smagorinsky model, and the new dynamic mixed model is quite good across all metrics shown. Indeed, mixed models of this form have been demonstrated to have excellent physical fidelity and performance in LES of canonical flows \citep{Vreman1997,Portwood2021}. 
%A priori testing is shown here not as proof of a good model, as only a posteriori testing can demonstrate, but only as a first step in assessing the first-fruits of the physics-inspired coarsening framework.
%The physics-inspired approach offers a more robust framework for a posteriori testing of these models and the likely development of further improvements.

\subsection{\textit{A posteriori} testing}

Large eddy simulations were performed using three stress models: Smagorinsky, PIC-based dynamic eddy viscosity (Eq.\ \eqref{eq:dyn-nut}), and PIC-based dynamic mixed (Eq.\ \eqref{eq:dyn-mixed-model}). The same pseudo-spectral code from the DNS is used. The Smagorinsky coefficient is not determined dynamically for the LES, but was manually chosen to produce accurate results. For the two PIC-based dynamic models, spatial averaging is used for the numerator and denominator of the eddy viscosity, so that the eddy viscosity does not vary in space for either. As such, the simulation with the PIC-based dynamic eddy viscosity model is essentially a DNS at a lower Reynolds number, with the PIC-based theory setting the viscosity so as to match Gaussian-filtered DNS with a given filter width.

\begin{figure}
	\centering
	\includegraphics[width=0.49\linewidth]{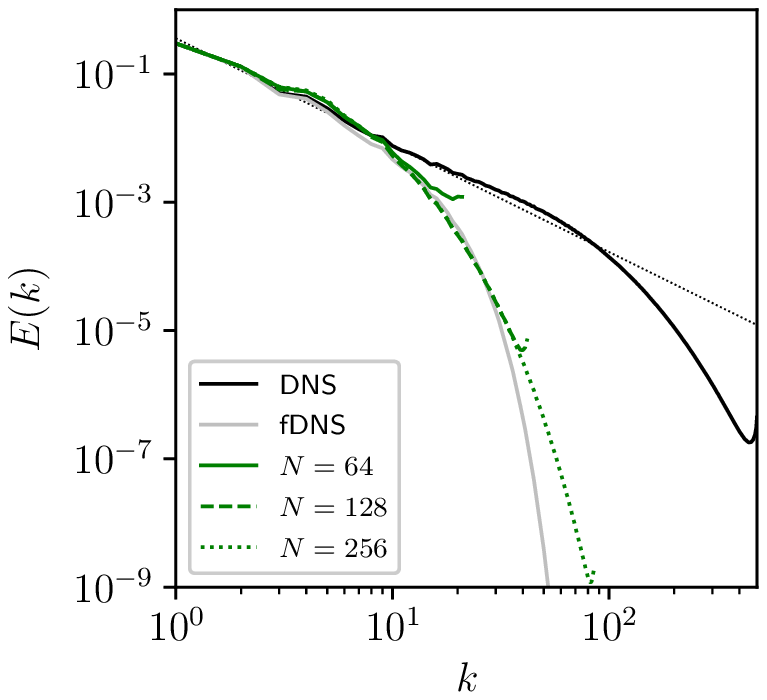}
	\includegraphics[width=0.49\linewidth]{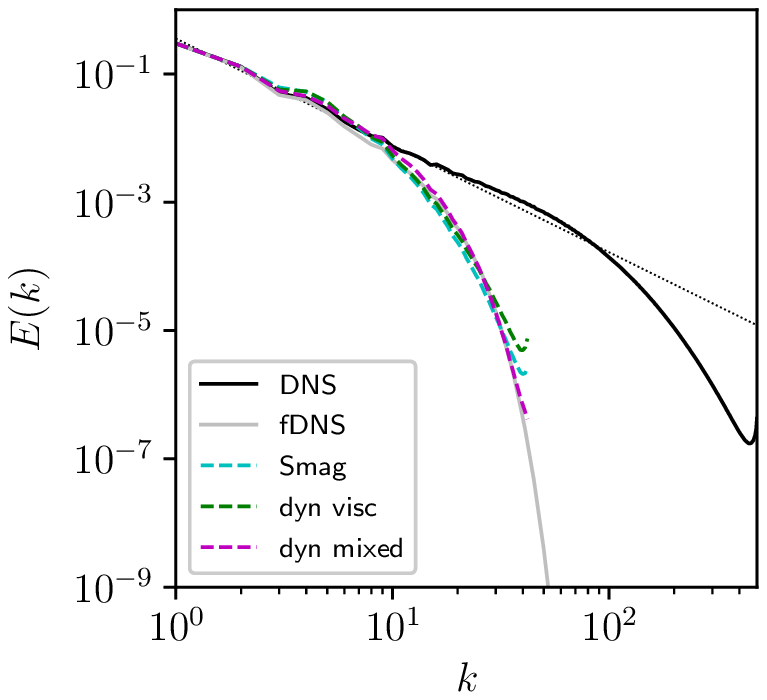}
	\caption{Energy spectra for \textit{a posteriori} testing: (a) using PIC-based dynamic eddy viscosity model with $\ell / \eta = 24$ at three different resolutions, and (b) using three different models with $k_{\text{max}} \ell = 3$. A Kolmogorov spectrum, $E(k)  =1.6 \epsilon^{2/3} k^{-5/3}$ is shown for reference in both panels.}
	\label{fig:LES-spectra}
\end{figure}

The filter width (pseudo-time) is chosen to be $\ell / \eta = 24$, which is at the small-scale end of the inertial range identified in the \textit{a priori} tests. The following grid resolutions were tested: $64^3$, $128^3$, and $256^3$. These correspond to $k_{\text{max}} \ell = 1.5$, $3.0$, and $6.0$. The DNS results are filtered with a Gaussian kernel (equivalent to PIC) for direct comparison with the LES results from each model. Figure \ref{fig:LES-spectra}a shows the energy spectra from LES simulations at each of the three resultions with the PIC-based dynamic eddy viscosity model. The spectra from each resolution overlap until wavenumbers close to $k_{\text{max}}$, where a slight pileup occurs. The results from $128^3$ and $256^3$ are very similar, indicating grid convergence, but $64^3$ is not grid-converged. The spectra from the three models are compared at $128^3$ resolution in Figure \ref{fig:LES-spectra}b. The Smagorinsky model is known to produce a spectrum close to that of a Gaussian filter \cite{Pope2000}, as affirmed by the results here. The PIC-based dynamic mixed model produces a spectrum almost indistinguishable from the Gaussian filtered DNS for the simulation shown here. All three models produce spectra in reasonable agreement. The PIC-based dynamic procedure for determining the eddy viscosity is thus shown to work well for both the pure eddy viscosity and mixed models.

In addition to testing the energy spectra produced by various LES models, it is also useful to compare local flow topology statistics. Given the importance of strain self-amplification and vorticity stretching to turbulence dynamics, the LES models are compared to filtered DNS in terms of the following two quantities,
\begin{equation}
	s^* = \frac{-\sqrt{6} S_{ij} S_{jk} S_{ki}}{\left(S_{mn} S_{mn}\right)^{3/2}}
	\hspace{0.1\linewidth}
	\omega^* = \frac{\sqrt{6} W_i S_{ij} W_j}{2 W_k W_k \left( S_{mn} S_{mn} \right)^{1/2}}.
	\label{eq:star}
\end{equation}
The first, $s^*$, was introduced by \citet{Lund1994} and quantifies the efficiency of the strain self-amplification contribution to the energy cascade, $-1 \leq s^* = \Pi_{s1} / \Pi_{s1,\text{max}} \leq 1$. The second term likewise indicates the efficiency of the vorticity stretching contribution to the cascade, $-1 \leq \omega^* = \Pi_{\omega 1} / \Pi_{\omega 1,\text{max}} \leq 1$. The concept of cascade efficiency was introduced by \citet{Ballouz2018} and extended to include the above definitions in \citet{Johnson2021b}. These two quantities describe flow topology in a way most relevant to turbulent cascade physics.

\begin{figure}
	\includegraphics[width=0.5\linewidth]{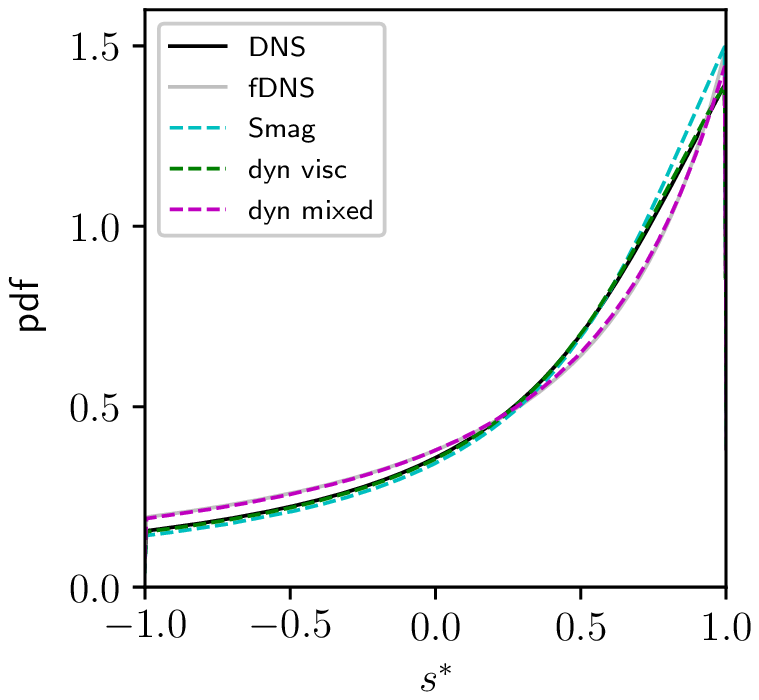}
	\includegraphics[width=0.5\linewidth]{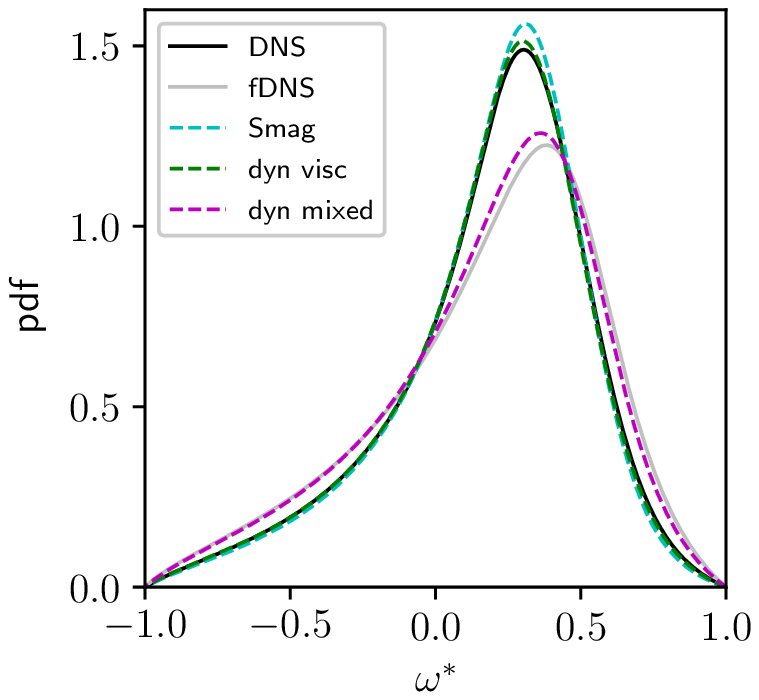}
	\caption{Probability density functions (PDFs) of $s^*$ and $\omega^*$, Eq.\ \eqref{eq:star}, for LES with three different models compared to DNS and filtered DNS.}
	\label{fig:LES-star}
\end{figure}

Figure \ref{fig:LES-star} compares the PDF of $s^*$ and $\omega^*$ from LES models and filtered DNS. Results from unfiltered DNS are also shown to highlight the similarity of the eddy viscosity models, particularly the PIC-based dynamic viscosity model, with (unfiltered) DNS. The reason for this is that the eddy viscosity models behave like DNS at lower Reynolds numbers, and thus produce topology statistics similar to the Kolmogorov scales in DNS. However, the results in Figure \ref{fig:LES-star} demonstrate that the PIC-based mixed model captures subtle physical differences between viscous-scale and inertial range flow topology statistics.

Many other metrics can be used to judge the \textit{a posteriori} accuracy of LES models. It is not the goal at present to thoroughly explore all possibilities. However, a number of highly relevant statistics are summarized in Table \ref{tab:LES-compare}. It may be noted that the PIC-based dynamic mixed model slightly over-predicts the magnitude of velocity gradients, which may also be noticed in the spectra shown in Figure \ref{fig:LES-spectra}b, where a slight over-prediction is noticeable for a range of wavenumbers slightly smaller than the inverse of the filter scale. There appears to be a slight (unphysical) build up of energy not entirely unlike the bottleneck effect commonly observed in DNS and experimental data near the Kolmogorov scale \citep{Falkovich1994, Donzis2010}.
	
All three models over-predict the fraction of the cascade rate due to single-scale strain self-amplification and vorticity stretching. For the mixed model, this shows that too much energy removal is done through the nonlinear gradient term, and not enough through the eddy viscosity. The longitudinal and transverse velocity gradient skewness and flatness values from the mixed model LES are in fairly good agreement with the filtered DNS. Notice that the dynamic eddy viscosity model leads to too much intermittency (higher skewness and flatness), because it has a spatially uniform eddy viscosity, whereas the local adjustments due to Smagorinsky's eddy viscosity bring intermittency metrics more in line with filtered DNS.

Overall, all three models perform reasonably well in \textit{a posteriori} tests. Furthermore, the relative advantages of the mixed model are clear, and the success of the PIC-based dynamic scheme is demonstrated. It should be appreciated that PIC theory for LES can facilitate model development well beyond the specific models tested here. The success of these models, however, does motivate future work to develop refined PIC-based models.

\begin{table}
	\centering
	\begin{tabular}{ccccccc}
		model & $\langle S_{ij} S_{ij} \rangle \tau_\eta^2$ & $\langle \Pi_{s1} \rangle / \langle \Pi \rangle$ & $\langle \Pi_{\omega 1} \rangle / \langle \Pi \rangle$ & $\langle A_{11}^3 \rangle / \langle A_{11}^2 \rangle^{3/2}$ & $\langle A_{11}^4 \rangle / \langle A_{11}^2 \rangle^{2}$ & $\langle A_{12}^4 \rangle / \langle A_{12}^2 \rangle^{2}$ \\
		\hline
		fDNS & $1.7$e-$2$ & $0.37$ & $0.12$ & $-0.41$ & $3.9$ & $4.6$ \\
		Smag. & $1.8$e-$2$ & $0.42$ & $0.14$ & $-0.42$ & $3.6$  & $4.9$ \\
		dyn. visc. & $1.9$e-$2$ & $0.50$ & $0.17$ & $-0.47$ & $4.7$ & $6.7$ \\
		dyn. mixed & $1.9$e-$2$ & $0.45$ & $0.15$ & $-0.41$ & $4.0$ & $4.9$\\
		%\hline
	\end{tabular}
	\caption{Statistical results from LES and filtered DNS at $\ell / \eta = 24$ with $128^3$ resolution ($k_{\text{max}} \ell = 3$). These values were converged when compared with $256^3$ resolution.}
	\label{tab:LES-compare}
\end{table}

\section{PIC for more complex flows}\label{sec:complex}

In section \ref{sec:wall-free}, it was shown that physics-inspired coarsening (PIC) is mathematically equivalent to spatial filtering with a Gaussian kernel for unbounded flows with uniform, isotropic resolution. Even so, the PIC-based approach to LES provided theoretical and modeling insights, including an PIC-based alternative to the Germano-based dynamic procedure that does not require test filtering. Preliminary models showed promising \textit{a priori} and \textit{a posteriori} results, but there is more room for future work developing and testing models even for that simple case.

In this section, the extension of PIC to a number of more complex flow scenarios is outlined. The focus is on how the basic theory of PIC provides the flexibility and systematic approach to incorporate features needed for practical application of LES to complex flows.

\subsection{Anisotropic resolution}

The first step toward LES of more complex flows is a consistent treatment of grid anisotropy. Anisotropic grid resolutions are often unavoidable in simulations with complex geometries. Various treatments of grid anisotropy exist for spatial filtering theory \citep{Bardina1980, Scotti1993, Vreman2004, Rozema2015, Haering2019}. In practice, simple treatments of grid anisotropy are common, such as using a scalar measure for effective resolution based on a suitable average of resolution in three different directions on an orthogonal grid: $\Delta_1$, $\Delta_2$, and $\Delta_3$. \citet{Deardorff1970} suggested the use of $\Delta = \sqrt[3]{\Delta_1 \Delta_2 \Delta_3}$ based on the cell volume and \citet{Bardina1980} used $\Delta = \sqrt{(\Delta_1^2 + \Delta_2^2 + \Delta_3^2)/3}$.

\begin{figure}
	\includegraphics[width=1.0\linewidth]{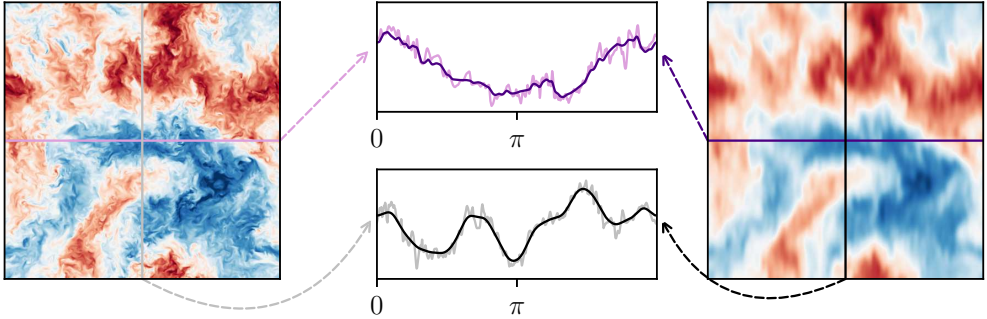}
	\caption{A coarsened velocity field using anisotropic pseudo-viscosity on DNS data of homogeneous isotropic turbulence. The resolution length scale ($\sqrt{2\hat{\nu}\hat{t}}$) in the x direction (purple/magenta lines) is one-fourth of that in the y direction (black/gray lines).}
	\label{fig:anisotropic}
\end{figure}

For viscosity-based smoothing, anisotropic resolution is easily introduced using an tensorial pseudo-viscosity aligned with the principle directions of the grid.
Thus, Eq. \eqref{eq:aux-Stokes-free} for unbounded flows may be modified to incorporate anisotropic resolution effects,
\begin{equation}
	%\frac{\partial \overline{u}_i}{\partial \hat{t}} + \frac{\partial}{\partial x_j}\left( \hat{p}\delta_{ij} - \hat{\nu}_{jk} \frac{\partial \overline{u}_i}{\partial x_k} + \hat{\nu}_{ik} \frac{\partial \overline{u}_j}{\partial x_k} \right) = 0,
	\frac{\partial w_i}{\partial \hat{t}} = \hat{\nu}_{jk} \frac{\partial^2 w_i}{\partial x_j \partial x_k},
	\hspace{0.1\linewidth}
	\frac{\partial w_j}{\partial x_j} = 0.
	\label{eq:aux-Stokes-anisotropic}
\end{equation}
For uniform resolution, the pseudo-pressure will satisfy the Laplace equation and hence be $\hat{p} = 0$ in the absence of flow boundaries. In that case, the result will be anisotropic Gaussian filtering aligned with the grid,
\begin{equation}
	w_i(\mathbf{x}, t; \hat{t}) = \iiint \frac{1}{\sqrt{ (2 \pi)^3 \det \mathbf{C}  } } \exp\left( - \frac{1}{2} r_j ~C_{jk}^{-1} r_k  \right) u_i(\mathbf{x}+\mathbf{r}, t) d\mathbf{r}.
\end{equation}
Here, the covariance tensor, $C_{ij} = 2 \hat{\nu}_{ij} \hat{t}$, is equivalent to the moment of inertial tensor of \citet{Bardina1980}. The isotropic form, Eq.\ \eqref{eq:aux-Stokes-free}, is recovered when $\hat{\nu}_{ij} = \hat{\nu} \delta_{ij}$. The eigenvalues of $\mathbf{C}$, $\ell_{1,2,3}^2 = 2 \hat{\nu}_{1,2,3} ~\hat{t}$, represent the (square of the) resolution in three directions given by their respective eigenvectors (e.g., the local coordinate frame of an orthogonal grid).
The result of anisotropic PIC (i.e., anisotropic Gaussian filtering) is illustrated in Figure \ref{fig:anisotropic}. In this case, the coarsening procedure generates anisotropy in the generalized velocity field from an initially isotropic fully-resolved velocity field \citep{Haering2019}.

Substituting Eqs. \eqref{eq:gen-Navier-Stokes} and \eqref{eq:aux-Stokes-anisotropic} in Eq.\ \eqref{eq:mixed-partial}, one obtains,
\begin{equation}
	\frac{\partial \sigma_{ij}}{\partial \hat{t}} = \hat{\nu}_{mn} \frac{\partial^2 \sigma_{ij}}{\partial x_m \partial x_n} + \hat{\nu}_{mn} \left( \frac{\partial w_i}{\partial x_m} \frac{\partial w_j}{\partial x_n} + \frac{\partial w_i}{\partial x_n} \frac{\partial w_j}{\partial x_m} \right).
	\label{eq:residual-stress-diffusion-anisotropic}
\end{equation}
The details of the derivation leading to Eq.\ \eqref{eq:residual-stress-diffusion-anisotropic} follow the same steps as those shown in section \ref{sec:wall-free} leading to Eq.\ \eqref{eq:residual-stress-diffusion}. Indeed, it is readily verified that Eq.\ \eqref{eq:residual-stress-diffusion} is recovered in the isotropic limit, $\hat{\nu}_{ij} = \hat{\nu} \delta_{ij}$.
The PIC-based dynamic procedure for anisotropic grids can be based on Eq.\ \eqref{eq:residual-stress-diffusion-anisotropic}, which could lead to eddy viscosity models that share some similarities with the anisotropic minimum dissipation model \citep{Rozema2015} and the M43 model \citep{Haering2019}, while a PIC-based dynamic mixed model may share similarities with that of \citet{Vreman2004}.

%Make $\hat{\nu}$ into an anisotropic tensor reflecting the local grid topology.

\subsection{Nonuniform resolution}
\label{sec:nonuniform}
%Re-derive governing equation for residual stress, but with spatially varying pseudo-viscosity, $\hat{\nu} = \hat{\nu}(\mathbf{x})$. The pseudo-pressure, $\hat{p}$, will be proportional to derivatives of the pseudo-viscosity, enforcing divergence-free condition and avoiding extra commutator terms even in the presence of nonuniform resolution. The symmetry of mixed partial derivatives, Eq. \eqref{eq:mixed-partial}, may again be used to derive a PDE for the residual stress tensor.

Uniform grid resolution is useful for some simple canonical turbulent flows, but it is not practical for many naturally-occurring and engineered flows. As discussed in the introduction, a non-uniform filter does not commute with spatial differentiation, giving rise to additional terms in the governing equations for filtered fields \citep{Ghosal1995, Yalla2021}. Most notably, the filtered velocity field is no longer divergence-free \citep{Langford2001}.

In contrast, the physics-inspired coarsening (PIC) approach to the LES equations explicitly enforces a divergence-free condition on the generalized velocity field, $\mathbf{w}$, using a pseudo-pressure, $\hat{p}$, in the auxiliary evolution equation in pseudo-time, $\hat{t}$.
The above theory for uniform resolution was able to ignore the proposed pseudo-pressure  due to the form of its Poisson equation, Eq.\ \eqref{eq:pseudo-Poisson}, which may be alternatively written in terms of pseudo-viscosity gradients,
\begin{equation}
	\nabla^2 \hat{p} = 2 \frac{\partial \hat{\nu}}{\partial x_i} \nabla^2 w_i + \frac{\partial^2 \hat{\nu}}{\partial x_i \partial x_j}\left( \frac{\partial w_i}{\partial x_j} + \frac{\partial w_j}{\partial x_i} \right).
	\label{eq:pseudo-Poisson-nut}
\end{equation}
which is a Laplace equation for uniform pseudo-viscosity (i.e., uniform grid resolution).
Nonuniform resolution is represented as spatial variation of the pseudo-viscosity, $\hat{\nu} = \hat{\nu}(\mathbf{x})$, which activates a non-zero pseudo-pressure to enforce the divergence free condition.
Note that, with the nonzero pseudo-pressure, the PIC equations for pseudo-time evolution are no longer purely parabolic, but include an elliptic nature as well. However, the square-of-exponential falloff of the spectral content is still roughly preserved, so that the removal of small-scales is still rather efficient compared to other elliptic PDE based filters \citep{Germano1986a, Germano1986b, Bull2016}.

\begin{figure}
	\includegraphics[width=1.0\linewidth]{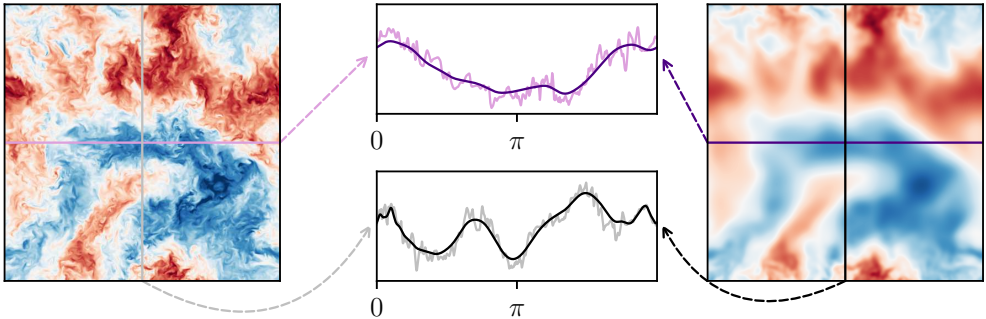}
	\caption{A (divergence-free) coarsened velocity field using sinusoidally-varying pseudo-viscosity applied to DNS data of homogeneous isotropic turbulence. The resolution length scale, $\sqrt{2\hat{\nu}\hat{t}}$ varies by a factor of $4$ between $y = 0$ and $y = \pi$.}
	\label{fig:nonuniform}
\end{figure}

Figure \ref{fig:nonuniform} demonstrates the outcome of PIC with a spatially-varying pseudo-viscosity for a velocity fied snapshot from DNS of homogeneous isotropic turbulence with $\hat{\nu}(y) = \hat{\nu}_{\max} - (\hat{\nu}_{\max} - \hat{\nu}_{\min}) \cos(y)$, where $0 \leq y \leq 2\pi$ is the vertical axis as printed on the page. The rightmost image is of a divergenge-free velocity field with variable resolution.
The effective resolution is isotropic, but varies in space, with high resolution near the top and bottom ($\hat{\nu} \approx \hat{\nu}_{\min}$) and low resolution near the center ($\hat{\nu} \approx \hat{\nu}_{\max}$).

The case of nonuniform (isotropic) resolution is useful for reflecting on the physics-inspired coarsening (PIC) approach in general. The goal of PIC is to create LES equations with well-defined closure terms that facilitate effective modeling in ways that improve upon spatial filtering. It is not a central goal of PIC to remain particularly faithful to physical processes during its pseudo-time evolution. The physical processes inspiring the pseudo-time dynamics are useful insomuch as they create helpful representations for LES. This allows some freedom when designing the details of PIC, for example, one may choose to simplify the auxiliary Stokes equation, Eq.\ \eqref{eq:aux-Stokes}, to,
\begin{equation}
	\frac{\partial w_i}{\partial \hat{t}} = -\frac{\partial \hat{p}}{\partial x_i} + \frac{\partial}{\partial x_j}\left[ \hat{\nu} \frac{\partial w_i}{\partial x_j} \right],
	\hspace{0.05\linewidth}
	\frac{\partial w_j}{\partial x_j} = 0,
	\hspace{0.05\linewidth}
	\nabla^2 \hat{p} = \frac{\partial}{\partial x_j}\left( \frac{\partial \hat{\nu}}{\partial x_i} \frac{\partial w_i}{\partial x_j} \right).
	\label{eq:aux-Stokes-nonuniform}
\end{equation}
The difference between Eq.\ \eqref{eq:aux-Stokes} and Eq.\ \eqref{eq:aux-Stokes-nonuniform} is only material for nonuniform pseudo-viscosity. Either may be used for PIC, and it is not immediately evident which will be most beneficial for LES. More research is needed to elucidate the relative strengths of either.

Substituting Eqs. \eqref{eq:aux-Stokes-nonuniform} and \eqref{eq:gen-Navier-Stokes} into Eq. \eqref{eq:mixed-partial} and performing simplifications like those in \S\ref{sec:wall-free}, one may arrive at the auxiliary equation for the residual stress for general $\hat{\nu} = \hat{\nu}(\mathbf{x})$,
%\begin{multline}
%	\frac{\partial \tau_{ij}}{\partial \hat{t}} =
%	\frac{\partial}{\partial x_k}\left( \hat{\nu} \frac{\partial \tau_{ij}}{\partial x_k}  \right)
%	+ 2 \hat{\nu}  \frac{\partial \overline{u}_i }{\partial x_k} \frac{\partial \overline{u}_j}{\partial x_k}
%	- \left(
%	2 \overline{u}_i \frac{\partial \overline{u}_k}{\partial x_j}
%	+ 2 \overline{u}_k \overline{S}_{ij}
%	+ \frac{\partial \tau_{ik}}{\partial x_j}
%	- 4 \nu \frac{\partial \overline{S}_{ij}}{\partial x_k}
%	\right) \frac{\partial \hat{\nu}}{\partial x_k}
%	\\
%	-  \left(
%	\overline{u}_k \frac{\partial \overline{u}_j}{\partial x_k}
%	+ \frac{\partial \tau_{jk}}{\partial x_k}
%	+ 2 \frac{\partial \overline{p}}{\partial x_j}
%	\right) \frac{\partial \hat{\nu}}{\partial x_i}
%	+ 2 \nu \overline{S}_{ij} \nabla^2 \hat{\nu}
%	+ \overline{u}_i \frac{\partial \hat{p}}{\partial x_j}
%	+ \overline{u}_j \frac{\partial \hat{p}}{\partial x_i}
%	\label{eq:residual-stress-diffusion-nonuniform}
%\end{multline}
\begin{multline}
	\frac{\partial \sigma_{ij}}{\partial \hat{t}} = 
	\frac{\partial}{\partial x_k}\left( \hat{\nu} \frac{\partial \sigma_{ij}}{\partial x_k} \right)
	+ 2 \hat{\nu}  \frac{\partial w_i }{\partial x_k} \frac{\partial w_j}{\partial x_k} \\
	- \left( w_i \frac{\partial w_k}{\partial x_j}
	+ w_j \frac{\partial w_k}{\partial x_i} + \frac{\partial w_i w_k}{\partial x_j} + \frac{\partial \sigma_{ik}}{\partial x_j}
	- \frac{\partial \hat{p}}{\partial x_j} \delta_{ik} \right) \frac{\partial \hat{\nu}}{\partial x_k}
	+ \left( w_i \delta_{jk} + w_j \delta_{ik} \right) \frac{\partial \hat{p}}{\partial x_k}.
	\label{eq:residual-stress-diffusion-nonuniform}
\end{multline}
Deviations from Eq. \eqref{eq:residual-stress-diffusion} are proportional to the pseudo-viscosity gradient, as well as the pseudo-pressure. One way to view the extra terms in Eq.\ \eqref{eq:residual-stress-diffusion-nonuniform} is to view them as consequences of the lack a commutivity between multiplication by $\hat{\nu}$ and differentiation. The advantage of this PIC formulation is that these commutator terms are built into the residual stress, and so provide a path for extending the use of dynamic procedures and other approaches from uniform resolution to more general cases.
Beyond traditional (theory-based) modeling approaches, data-driven techniques may stand to benefit even more from PIC and Eq.\ \eqref{eq:residual-stress-diffusion-nonuniform}.
While isotropic nonuniform resolution has been considered here, the same procedure may be used to combine anisotropic and nonuniform grid effects into a single formulation.

% reference to supplementary materials (detailed derivation) ??? -- only if they ask

\subsection{Heat and mass transfer}

In addition to complexities introduced by numerical grid effects, physics-inspired coarsening provides a holistic approach for reducing the computational DoFs needed to represent turbulence in regimes with additional physical effects beyond unbounded single-phase incompressible flows. The next few subsections touch briefly on a few examples, the simplest of which is passive scalar transport.

For flows with heat or mass transfer
described by an advection diffusion equation,
\begin{equation}
	\frac{\partial T}{\partial t} + u_j \frac{\partial T}{\partial x_j} = \alpha \nabla^2 T
\end{equation}
the relevant scalar field $T$ (e.g., temperature or concentration) may be coarsened in a physics-inspired manner using an pseudo-conductivity or pseudo-diffusivity, $\hat{\alpha}$, to form a generalized temperature/concentration, $\theta$, with pseudo-time evolution given by an auxiliary heat equation,
\begin{equation}
	\frac{\partial \theta}{\partial \hat{t}} = \hat{\alpha} \nabla^2 \theta,
	\hspace{0.05\linewidth}
	\frac{\partial \theta}{\partial t}
	+ w_j \frac{\partial\theta}{\partial x_j}
	= \alpha \nabla^2 \theta
	- \frac{\partial q_j}{\partial x_j}
\end{equation}
so that the physical time evolution of the generalized temperature/concentration must include a residual heat/mass flux $\mathbf{q}$. The equality of mixed partial derivatives for $\theta$,
\begin{equation}
	\frac{\partial}{\partial \hat{t}}\left( \frac{\partial \theta}{\partial t} \right) 
	= \frac{\partial}{\partial t}\left( \frac{\partial \theta}{\partial \hat{t}} \right)
\end{equation}	
may be satisfied by a forced diffusion equation for the residual heat/mass flux,
\begin{equation}
	\frac{\partial q_i}{\partial \hat{t}} =
	\hat{\alpha} \nabla^2 q_i
	+ 2 \hat{\alpha} \frac{\partial w_i}{\partial x_j} \frac{\partial \theta}{\partial x_j}.
	\label{eq:residual-heat-flux-diffusion}
\end{equation}
Here, $\hat{\alpha} = \hat{\nu}$ is chosen (unity pseudo-Prandtl number) for convenience. Other choices of pseudo-Prandtl number are possible, but unlikely to be advantageous.

Similar extensions to nonuniform or anisotropic resolution may be found for heat/mass transfer as shown for the momentum equation above. Equation \eqref{eq:residual-heat-flux-diffusion} can also provide insight into scale-wise scalar fluctuation dynamics. A turbulent cascade of scalar variance may be linked to the multiscale squeezing of scalar filaments using the formal solution in wall-free flows following the procedure outline for the kinetic energy cascade in section \ref{sec:wall-free}. Furthermore, Eq.\ \eqref{eq:residual-heat-flux-diffusion} can serve as the basis for a PIC-based dynamic procedure for an eddy diffusivity or mixed model for $\mathbf{q}$.

%For variable density flows, a diffusion equation for the product $\rho \mathbf{u}$ could be used to prohibit net mass flux across streamlines of the coarsened field, as in Favre-filtering. %A pseudo-pressure may not be necessary for variable density flows that do not satisfy a divergence-free condition.

\subsection{Wall-bounded turbulent flows}

Turbulence near solid boundaries is very common in flows of interest for LES. This is the source of some inconvenience for spatial filtering theory. A uniform filter integral will require flow information outside the fluid domain \citep{Drivas2018}. This may be avoided through the use of a filter width which vanishes at the boundary is approached \citep{Bose2014}, giving rise to commutation errors discussed in \S\ref{sec:nonuniform}.

In contrast to spatial filtering, the physics-inspired approach naturally extends to wall-bounded flows. To accomplish this, the auxiliary Stokes equation, Eq. \eqref{eq:aux-Stokes}
%or \eqref{eq:aux-Stokes-free}, 
must be given boundary conditions for $\mathbf{w}$.
For example, the use of user-defined boundary conditions for coarsening procedures has been preliminarily investigated \citep{Bae2017, Bae2018}. A number of choices are possible depending on the level of near-wall resolution. If the wall-normal grid spacing resolves the viscous sublayer, $\ell_y^+ \sim 1$, then the best choice of boundary conditions is likely Dirichlet, $\mathbf{w}|_{\text{wall}} = 0$. Depending on the wall parallel resolution, this approach could facilitate wall-resolved LES ($\ell_x^+ \sim \ell_z^+ \sim 50$) or near-wall RANS for hybrid RANS-LES (coarser x-z resolution).

Alternatively, a coarser wall-normal resolution, $\ell_y \gg 1$, would facilitate development of approximate closures for wall-modeled LES. In this case, Dirichlet boundary conditions could still be used \citep{Bae2021}, but other choices such as Robin \citep{Bose2014} or Neumann boundary conditions could be used. The choice of boundary conditions for the pseudo-time dynamics would directly set the boundary conditions to be used in LES on the generalized velocity field. Furthermore, complementary boundary conditions on the residual stress tensor, $\boldsymbol{\sigma}$, would also need to be specified in such a way as to preserve the integral of momentum transport across the boundary during the pseudo-time evolution so as to facilitate accurate calculations of forces in the LES. For example, in a wall-resolved LES regime, Dirichlet boundary conditions would likely be the best choice for the residual stress in conjunction with Dirichlet velocity boundary conditions. Further research is needed to determine the best approach for wall-modeled LES.

Data-driven techniques may be useful for testing the potential fruitfulness of various choices for the boundary conditions.
For example, the art of constructing data-driven wall models may benefit substantially from the PIC-based approach to the LES equations.
Sufficiently far from the wall, $y \gg \ell$, the physics-inspired coarsening naturally recovers Gaussian filtering, but near the wall, the PIC theory outlined in this paper provides a method for determining how the choice of boundary conditions influences the residual stress tensor.

\subsection{Particle-laden flows}

Turbulent flows with small particles also represent a challenge to spatial filtering theory. Lagrangian particle tracking methods, in combination with an Eulerian representation of the carrier fluid, are increasingly common to computing particle-laden flows. In the physics-inspired coarsening approach, the flow disturbance in the vicinity of the particle would be smoothed out, while boundary conditions may be freely chosen on the surface of the particle, as in the case of wall-bounded flows above. For example, choosing Neumann boundary conditions would, for particles smaller than the resolution length scale, $\ell \sim \sqrt{2 \hat{\nu} \hat{t}}$, cause the carrier flow velocity on the surface of the particle to relax to that of the far-field coarsened velocity. This would be similar to the idea of recovering the `undisturbed' fluid velocity upon which drag law formulations are based \citep{Horwitz2016, Horwitz2018, Balachandar2019} generalized to the case in which the Kolmogorov scale is not captured by the grid. The considerations for particle-resolved calculations would be the same as for wall-bounded turbulence the previous subsection.

\subsection{Multiphase flows with resolved interfaces}

Multiphase turbulent flows include the additional complexity of moving discontinuities at phase interfaces with the accompanying surface tension force. Indeed, the interface is ``a critical feature of such flows and it is likely that coarsening of the flow must retain the interface, although probably with a simplified structure'' \citep{Tryggvason2020}. 
Difficulties arise in spatial filtering across discontinuities \citep{Sagaut2005, Toutant2009}, and spatially filtering across fluid-fluid interfaces is not consistent with the typical (and desirable) sharp treatment of solid boundaries (walls) in LES.
Indeed, spatial filtering across interfaces blurs unnecessarily blurs essential physics at the interface, given the sophisticated numerical tools that have developed for treating sharp interfaces.

Various alternatives to spatial filtering for LES of multiphase flows have been proposed (see \citet{Cheng2019} for a more thorough, up-to-date review).
\citet{Herrmann2013} proposed a dual-scale resolution procedure that requires fine resolution of the interface embedded within coarser flow resolution.  \citet{McCaslin2014} suggested, but did not pursue, the idea of a surface filtering operation. Most recently, \citet{Chen2021} has advocated for a procedure very similar to the physics-inspired approach outlined here. Indeed, their approach, which specifically targets the use of data-driven modeling, could be thought of as one possibility within the PIC framework.

\begin{figure}
	\includegraphics[width=1.0\linewidth]{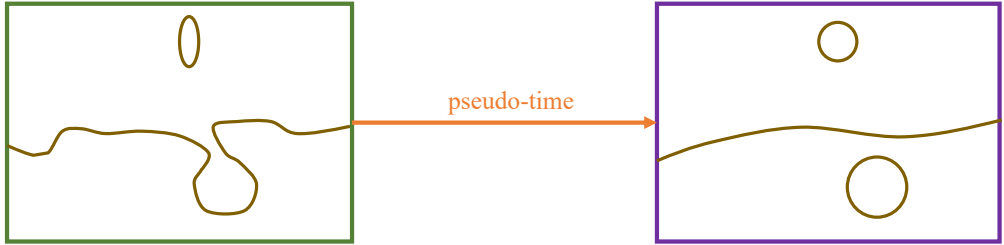}
	\caption{Sketch of how a pseudo surface tension could create low DoF representations of a complex two-phase interface.}
	\label{fig:surface-tension}
\end{figure}

One possibility suggested by the physics-inspired philosophy is the use of an artificial surface tension along with the auxiliary Stokes equation to remove small-scale features. Physically speaking, surface tension limits the impact of turbulent motions on multiphase features (drops, bubbles, etc) smaller than the Hinze scale \citep{Hinze1955}. Thus, just as the pseudo-viscosity artificially enhances the Kolmogorov scale, a pseudo surface tension acting during the pseudo-time evolution would likewise smooth small scale interface curvature while transforming more abrupt small-scale features into spherical particles for Lagrangian particle tracking, see Figure \ref{fig:surface-tension}. Note that this PIC process would maintain a sharp interface treatment for features that can be captured on the coarse LES grid while creating low DoF representations for unresolved features. Such an approach provides a general framework to justify other existing heuristic approaches, e.g., for physics-based conversion of subgrid ligaments to point-particle representations \citep{Kim2020}. The use of an artificial diffusion equation for Gaussian filtering has some precedence for simulating particle-laden flows \citep{Capecelatro2013}. An artificial surface tension may also be a useful approach for treating multiscale (partially-resolved) wall roughness.

\section{Conclusions}\label{sec:conclusions}

In turbulence, viscosity provides a natural mechanism preventing motions smaller than the Kolmogorov scale. In this paper, it is proposed to view an LES velocity field as the result of (artificial) viscous smoothing rather than a spatial filtering operating. This approach is called physics-inspired coarsening (PIC). In PIC, a pseudo-viscosity (and pseudo-pressure) acts in pseudo-time on an initially fully-resolved snapshot of turbulence according to an auxiliary Stokes equation. The generalized velocity field is a function of both physical time and pseudo-time. The effective equations for the generalized velocity field in physical time (at a fixed pseudo-time corresponding to numerical resolution $\ell \sim \sqrt{2 \hat{\nu} \hat{t}}$) provide the governing equations for LES. The equality of mixed partial derivatives in physical time and pseudo-time provides a consistency condition that defines the residual stress tensor based on the chosen pseudo-time evolution equation.

In the simple case of unbounded flow with uniform numerical resolution, PIC is shown to be equivalent to Gaussian filtering. Thus, the advantages of spatial filtering are retained while providing a basis for defining a more generally applicable coarsening procedure for complex flows. Even in the simple case of unbounded flows with uniform resolution, the PIC approach yields fresh insight.
First, PIC leads to a previously-hidden exact representation of the energy cascade in terms of multiscale velocity gradient interactions. This provides a more precise framework for representing residual stresses based on vortex stretching and other similar processes.  Furthermore, the resulting parabolic PDE in pseudo-time may be used to construct a Germano-like dynamic procedure that does not require a test filter. This is demonstrated in the creation of a dynamic eddy viscosity and dynamic mixed model based on PIC equations. The performance of both models in \textit{a priori} and \textit{a posteriori} testing is demonstrated in homogeneous isotropic turbulence. The dynamic eddy viscosity model performs similarly to the commonly-used Smagnorinsky model and the relative advantages of the dynamic mixed model are illustrated.

For realistic flows where nonuniform grid resolution is desirable, the PIC framework for LES provides a nature-inspired mechanism for maintaining a divergence-free condition for incompressible flows while avoiding commutation errors (extra terms in the LES equations). In the case of wall-bounded flows, boundary conditions may be chosen, PIC provides a convenient definition of what it means to coarsen a flow in the vicinity of a domain boundary, and can facilitate various modes of near-wall resolution or modeling treatment. PIC theory is also easliy extended for a consistent treatment of grid anisotropy by using an anisotropic pseudo-viscosity.

Finally, it is suggested that the physics-inspired framework for LES extends more naturally to flows with more complex physics. A simple example of such an extension is heat/mass transfer, for which the PIC approach to momentum is straightforwardly copied. More significant extensions are also possible with PIC, such as particle-laden or (interface-resolved) multiphase flows.
While PIC provides theoretical developments that can aid model development for complex LES, it also provides an important foundation for enabling robust data-driven modeling approaches. Indeed, data-driven closure techniques may benefit significantly from the advantages of the PIC approach to LES theory, for example, the recovery of divergence-free coarsened velocity field for incompressible flows.

% data-driven techniques for turbulence closure modeling mature, the opportunity to employ such approaches in LES for complex multiphysics flows requires a framework for defining what a good LES closure model must represent. 

%Given its established role in LES theory, spatial filtering theory seems like a natural choice for this role. However, the physics-inspired framework introduced in this paper overcomes the intrinsic difficulties in the spatial filtering approach.

%Thus, this new perspective on LES theory gives provides a more promising basis for training and deploying data-driven modeling techniques. Physics-based closure modeling approaches, such as those concerned with coherent structures, will also benefit from this perspective.

\section*{Declaration of Interests}
The author reports no conflict of interest.

% Bibliography
\bibliographystyle{jfm}
\bibliography{picles}

\end{document}